\documentclass[druk]{appolb}
\usepackage{graphicx}
\usepackage{amsmath,amssymb}
\usepackage{slashed}
\usepackage[]{hyperref}
\renewcommand{\runhead}{}
\begin{document}
\title{Gluonic structure from instantons\thanks{Contribution to the
memorial volume dedicated to Dmitri I.\ Diakonov, Victor Yu.\ Petrov, and Maxim V.\ Polyakov (2025)}}
\author{C.~Weiss
\address{Theory Center, Jefferson Lab, Newport News, VA 23606, USA}}
\maketitle
\begin{abstract}
The instanton vacuum provides an effective description of chiral symmetry breaking by
local topological fluctuations of the gauge fields, as observed in lattice QCD simulations.
The resulting effective dynamics at momenta below $1/\bar\rho \approx$ 0.6 GeV explains the basic
features of light-quark correlation functions and is used extensively in studies
of hadron structure. The instanton fields also make definite contributions to the
gluonic structure of light hadrons, as expressed in the matrix elements of composite quark-gluon
or gluon operators. The article reviews the gluonic structure of light hadrons (nucleon, pion)
induced by instantons.
This includes:
(i)~twist-2 parton distributions and momentum sum rule;
(ii)~twist-3 angular momentum and spin-orbit interactions;
(iii)~twist-3 and 4 quark-gluon correlations and power corrections;
(iv)~trace anomaly and hadron mass decomposition;
(v)~scalar gluon form factors and mechanical properties;
(vi)~axial anomaly and pseudoscalar gluon form factors.
It also discusses possible further applications of the methods and recent developments
including gauge field configurations beyond instantons.
\end{abstract}
{\small
\vspace{-4ex}
\tableofcontents
}
\section{Introduction}
\label{sec:intro}
Chiral symmetry breaking (ChSB) plays an essential role in the emergence of hadron structure from QCD.
It is connected with the dynamical generation of mass in the world of light hadrons,
including the baryons, and determines the effective dynamics governing their structure.
It gives rise to nearly massless bosonic excitations, the pions, and restricts the form
of their interactions with other hadrons. The long-distance behavior of strong interactions
on the scale $1/M_\pi$ can be described by an effective field theory based on ChSB.

ChSB in QCD is caused by topological fluctuations of the gauge fields. In the sense of
real-time evolution, these fields describe tunneling trajectories between configurations
in sectors with different winding number \cite{Callan:1976je,Callan:1977gz,Vainshtein:1981wh}.
The topological gauge fields induce zero-virtuality modes
of the fermion field with definite chirality. ChSB arises from the delocalization
of the zero modes at finite density of the topological gauge fields.
The direct connection between ChSB and the zero modes is apparent from the Banks-Casher theorem,
which states that the chiral condensate is proportional to the spectral density of the
Dirac operator at zero virtuality \cite{Banks:1979yr}.

%
%
\begin{figure}[b]
\parbox[c]{.5\columnwidth}{
\includegraphics[width=.5\columnwidth]{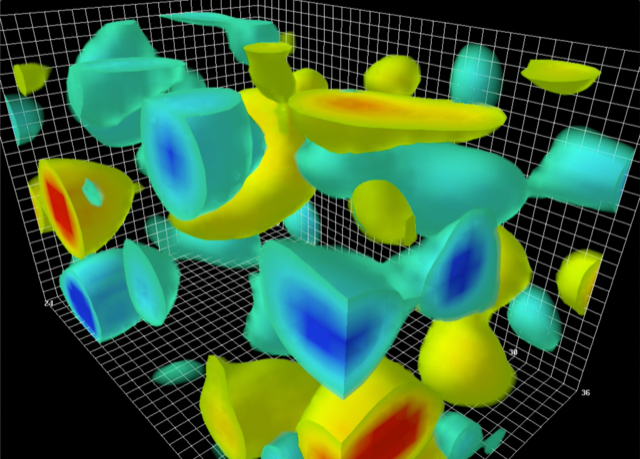}
}
\hspace{.05\columnwidth}
\parbox[c]{.4\columnwidth}{
\caption{Local concentrations of topological charge in cooled lattice
configurations of gluodynamics \cite{Leinweber:1999cw}. Yellow: Positive charge (instantons).
Blue: Negative charge (antiinstantons).}
}
\label{fig:cool}
\end{figure}
The quantitative features of the topological gauge field fluctuations and ChSB
in the imaginary-time (Euclidean) formulation have been investigated in lattice QCD.
Cooling methods suppress quantum fluctuations and produce smooth configurations showing
local concentrations of topological charge and action (see Fig.~\ref{fig:cool} for a visualization
\cite{Leinweber:1999cw}); similar results are obtained with modern gradient flow
techniques \cite{Bonati:2014tqa,Alexandrou:2017hqw,Athenodorou:2018jwu}.
The typical size of the topological fluctuations is $\bar\rho \sim$ 0.3 fm, much smaller than
the hadronic size $\sim$ 1 fm.
The typical distance between the topological fluctuations is $\bar R \sim$ 1 fm,
and only a small fraction of 4-dimensional Euclidean space is occupied by such fields,
$\pi^2 \bar\rho^4 / \bar R^4 \sim 0.1$.
The average field strength inside the topological fluctuations is
$(F_{\mu\nu} F_{\mu\nu})^{1/2} \sim (32 \pi^2 / \pi^2 \bar\rho^4)^{1/2}
\sim$ 2 GeV$^2$, which is very large on the hadronic scale.
Such strong fields present favorable conditions for a semiclassical
description \cite{Callan:1976je,Callan:1977gz}.

\begin{figure}[t]
\begin{tabular}{ll}
\includegraphics[width=.3\columnwidth]{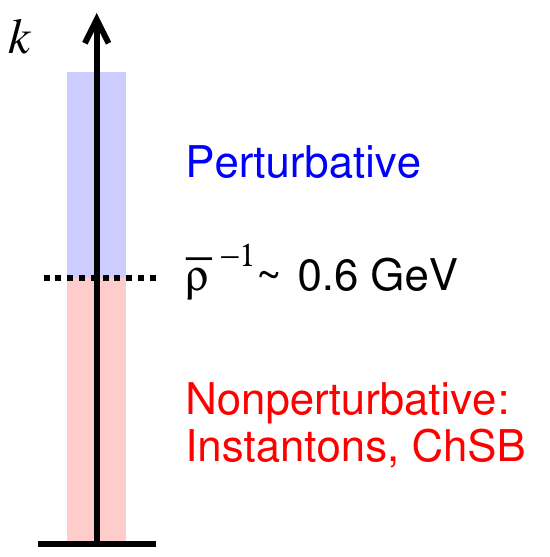}
\hspace{.1\columnwidth}
&
\includegraphics[width=.48\columnwidth]{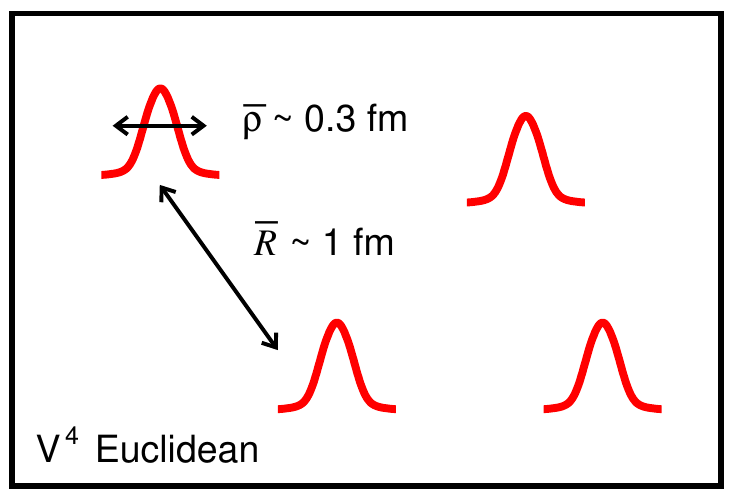}
\\[1ex]
{\small (a)} & {\small (b)}
\end{tabular}
\caption{Instanton vacuum. (a)~Separation of modes according to the dynamical scale $\bar\rho^{-1}$.
(b)~Instanton ensemble describing the low-momentum modes.}
\label{fig:instanton_vacuum}
\end{figure}

An effective description of ChSB by topological gauge fields is provided by the instanton vacuum
\cite{Shuryak:1981ff,Diakonov:1983hh}; see Refs.~\cite{Diakonov:2002fq,Schafer:1996wv} for reviews.
The basic idea is to separate the modes of the gauge fields according to the scale $\bar\rho^{-1}$ and
perform the functional integration using different approximations (see Fig.~\ref{fig:instanton_vacuum}).
The modes with momenta $k < \bar\rho^{-1}$ are described as a superposition of instantons
and instantons -- localized classical solutions with topological charge $\pm 1$, and
integrated using nonperturbative methods of statistical mechanics.
The modes with $k > \bar\rho^{-1}$ are integrated perturbatively and enter in the statistical weight
of the classical fields.
The procedure can be formulated as a variational approximation to the QCD partition
function \cite{Diakonov:1983hh}; questions such as the gauge dependence in the separation
of modes, ansatz dependence in the superposition of instantons, etc., are part of the ``choice
of trial function'' and contained in the overall variational approximation.
The construction uses the instanton packing fraction $\pi^2 \bar\rho^4 / \bar R^4 \ll 1$
as a small parameter \cite{Shuryak:1981ff} and employs it in the functional integration.
A stable ensemble is obtained by including instanton interactions derived from QCD \cite{Diakonov:1983hh}. The
quantitative features agree well with those observed in lattice QCD. The picture is robust
and does not depend on the details of the variational approximation \cite{Schafer:1996wv}.
%
%

ChSB in the instanton vacuum has been studied extensively and is well understood
\cite{Shuryak:1982dp,Diakonov:1985eg}; see Refs.~\cite{Diakonov:2002fq,Schafer:1996wv} for reviews.
The instantons induce multifermion interactions between the quarks through the fermion zero modes.
The finite density of instantons in the vacuum leads to the formation of a chiral condensate,
and the quarks acquire a dynamical mass $M \sim$ 0.3--0.4 GeV, of the order of a typical constituent quark mass.
The effective dynamics can be constructed and solved systematically in the
$1/N_c$ expansion \cite{Diakonov:1985eg,Diakonov:1986aj,Pobylitsa:1989uq,Nowak:1989at,Diakonov:1995qy,Kacir:1996qn}.
The interactions can be bosonized
and take the form of massive quarks interacting with chiral meson fields.
The resulting field theory captures the effective dynamics at Euclidean momenta
$k < \bar\rho^{-1}$ and describes a wide range of structures and phenomena.
Correlation functions in the meson sector exhibit the quasi massless pion pole in the
pseudoscalar-isovector channel and the massive $\eta'$ pole in the isoscalar channel
\cite{Diakonov:1986aj,Nowak:1989at,Kacir:1996qn}; see Ref.~\cite{Schafer:1996wv} for a review of other channels.
Correlation functions in the baryon sector are characterized by a localized classical
chiral field (``soliton''), in which the quarks move in independent-particle orbits \cite{Diakonov:1987ty},
providing a specific realization of the mean-field picture of baryons in the
large-$N_c$ limit of QCD \cite{Witten:1979kh}; see Ref.~\cite{Christov:1995vm} for a review.

In the effective dynamics emerging from ChSB the instanton gauge fields are subsumed in the
massive quark/antiquark degrees of freedom. Hadronic matrix elements of QCD quark operators
(vector or axial vector current, scalar density) can be obtained from the effective dynamics
without explicit reference to instantons. The instanton vacuum also enables the computation
of hadronic matrix elements of quark-gluon or pure gluon QCD operators, normalized
at the scale $\bar\rho^{-1}$ \cite{Diakonov:1995qy,Kacir:1996qn,Liu:2024rdm}.
In this context the instanton gauge fields appear explicitly and give rise to a definite
``gluonic structure'' of the light hadrons. Exploring this structure is interesting
for several reasons:

(i)~Gluon operators in the instanton vacuum are evaluated in an expansion in the
instanton packing fraction. The small parameter enables a systematic analysis and
establishes a hierarchy in the matrix elements of gluon operators with different
quantum numbers (spin, twist).
(ii)~The gluonic structure induced by instantons is derived using the same approximations
as in the effective dynamics emerging from ChSB. This preserves the essential connections
between quark and gluon operators, e.g. the momentum sum rule for twist-2 operators,
or QCD equation-of-motion relations for higher-twist operators.
(iii)~The instanton fields are strong on the hadronic scale. In channels where
single instantons are allowed to contribute they likely represent the dominant
effect in low-energy gluonic structure.
(iv)~The gluon operators enable the direct demonstration of instanton effects in hadron structure.
The selection rules implied by the symmetries of the instanton field are very distinctive and
can be compared with observations.
(v)~The instanton vacuum preserves the renormalization properties of QCD and implements the
conformal (trace) anomaly through instanton density fluctuations. It enables study of the 
interplay of conformal and chiral symmetry breaking, which is essential for the mass
decomposition of light hadrons. The instanton vacuum also implements the $U(1)_A$ axial anomaly
through topological charge fluctuations and enables study of its expression in hadron structure.

This article reviews the gluonic structure of light hadrons in the instanton vacuum.
It covers the basic methods, established and recent results, and suggestions for future
developments and applications. The treatment is based on the variational formulation
of the instanton vacuum of Refs.~\cite{Diakonov:1983hh,Diakonov:1985eg}
and the effective operator method of Ref.~\cite{Diakonov:1995qy},
which permits systematic calculation and characterization of gluonic structure.

Section~\ref{sec:dynamics} introduces the elements of the instanton ensemble and ChSB
needed in the present review. Section~\ref{sec:effop} describes the effective operator
method for the study of gluonic structure. Section~\ref{sec:gluonic} reviews the main
results in gluonic structure, organized according to the type of QCD operator;
conclusions and suggestions for further studies are presented at the end of each subsection.
Section~\ref{sec:beyond} describes recent developments in extending the semiclassical
approximation beyond instantons.

Theoretical models of the nonperturbative gluonic structure of the nucleon and other light hadrons
are urgently needed for many problems of current interest, such as generalized parton distributions,
the energy-momen\-tum tensor (EMT) form factors and hadron mass decomposition (trace anomaly),
higher-twist effects in inclusive and exclusive scattering, heavy quark contributions to
nucleon observables, heavy quarkonium production at near-threshold energies,
hadronic CP violation, and other phenomena.
The instanton vacuum can classify and estimate the gluon matrix elements in a systematic
fashion and make essential contributions to these areas of study.

The methods and results presented here are based on the renowned work of D.~I.~Diakonov
and V.~Yu.~Petrov on the instanton vacuum and represent only one its many contributions
to the understanding of nonperturbative QCD and hadron structure.
The applications to gluonic structure reviewed here were in large parts developed by
M.~V.~Polyakov and represent only a small part of his extensive and profound impact on
modern hadronic physics. The effective operator method was proposed in a work by Diakonov,
Polyakov, and this author as junior collaborator. I~had the fortune to work under
the guidance of DPP and learn from them over an extended time and consider this the
greatest blessing of my scientific and intellectual life. The best way in which our
community can honor their memory is to move ahead with the same energy and enthusiasm,
keep up the intellectual standards to the best of our abilities, realize the potential
of the concepts and methods they developed, and pass them on to the next generation.

\section{Effective dynamics from instantons}
\label{sec:dynamics}
The basic elements of the instanton vacuum are described in Refs.~\cite{Diakonov:2002fq,Schafer:1996wv}.
QCD is considered in 4-dimensional Euclidean space-time, with coordinates $x_\mu, \mu = 1 - 4,$
and metric $x^2 \equiv \sum x_\mu^2 > 0$. The normalization volume $V$ is finite and taken to be large,
with densities such as $N/V$ remaining stable in the limit.

%
%
\begin{figure}[t]
\begin{tabular}{ll}
\hspace{.1\columnwidth}
\includegraphics[width=.2\columnwidth]{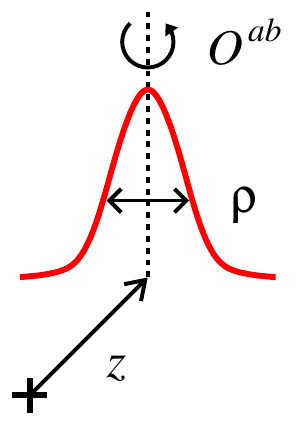}
\hspace{.1\columnwidth}
&
\includegraphics[width=.4\columnwidth]{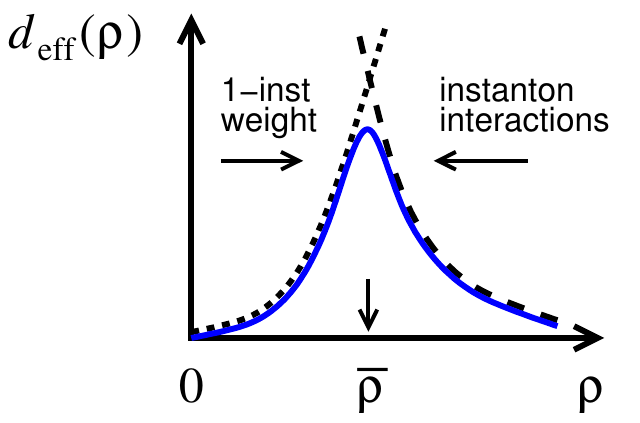}
\\[-2ex]
{\small (a)} & {\small (b)}
\end{tabular}
\caption{(a)~Collective coordinates characterizing the instanton.
(b)~Effective instanton size distribution Eq.~(\ref{d_eff}).}
\label{fig:collective_size}
\end{figure}
\textit{Instanton ensemble.}
The conceptual framework is a variational approximation to the gluodynamics
partition function \cite{Diakonov:1983hh}
(see comments in Sec.~\ref{sec:intro}). The low-momentum modes of the gauge potential ($k < \bar\rho^{-1}$)
are parametrized by a sum of instanton and antiinstanton potentials in singular gauge
(denoted by subscripts $\pm$),
\begin{align}
A(x) = \sum_{I}^{N_+} A_+ (x| z_I, O_I, \rho_I)
+ \sum_{\bar I}^{N_-} A_- (x| z_{\bar I}, O_{\bar I}, \rho_{\bar I});
\label{gauge_sum}
\end{align}
the explicit form of $A_\pm$ is given in Refs.~\cite{Diakonov:2002fq,Schafer:1996wv}.
Each instanton depends on a set of collective coordinates: the center coordinate $z$, color orientation
$O$, and size $\rho$ (see Fig.~\ref{fig:collective_size}a). The functional integration is performed as
\begin{align}
\int \prod_{I, \bar I}^{N_\pm} dz_I dO_I d\rho_I [...].
\end{align}
The high-momentum modes ($k > \bar\rho^{-1}$) are integrated out separately around each instanton,
as justified a posteriori by the diluteness of the instanton medium, and result in a statistical weight
per instanton \cite{tHooft:1976snw}
\begin{align}
d_0(\rho) = \textrm{const} \times \rho^{-5} (\rho \Lambda_{\rm QCD})^b \times \textrm{[NLO]},
\hspace{2em} b = \frac{11}{3} N_c - \frac{2}{3} N_f,
\label{d_0}
\end{align}
where $b$ is the LO coefficient of the QCD beta function; see
Refs.~\cite{Schafer:1996wv,Diakonov:2009jq} for details.
The weight Eq.~(\ref{d_0})
strongly increases at large sizes and would not in itself result in a stable system.

A stable system is obtained by including the effect of instanton interactions,
which suppress instantons with large size.
This can be done consistently and efficiently with the variational approximation of Ref.~\cite{Diakonov:1983hh}.
The trial partition function is chosen as an ensemble of independent instantons with an effective
size distribution,
\begin{align}
Z_{\rm int} = \int \prod_{I, \bar I}^{N_\pm} dz_I dO_I d\rho_I \; d_{\rm eff}(\rho_I) .
\label{trial}
\end{align}
Performing a variational estimate of the full QCD partition function in terms of the trial partition
function Eq.~(\ref{trial}),
one can naturally identify the instanton interactions generated by QCD and evaluate their effect
on the size distribution. The effective size distribution is obtained as 
\begin{align}
d_{\rm eff}(\rho) = \textrm{const} \times  d_0(\rho) e^{- \alpha\rho^2},
\label{d_eff}
\hspace{2em}
\alpha = \gamma \frac{8\pi^2}{g^2} \frac{N}{V} ,
\end{align}
where $\gamma$ is a constant characterizing the instanton interactions, $g$ is the
coupling constant at the scale $\bar\rho^{-1}$, and $N/V \equiv (N_+ + N_-)/V$ is the
total instanton density. The distribution Eq.~(\ref{d_eff}) suppresses large sizes and
leads to a stable system (see Fig.~\ref{fig:collective_size}b).
In the large-$N_c$ limit the width of $d_{\rm eff}(\rho)$
is $\mathcal{O}(1/N_c)$, so that fluctuations of the sizes are suppressed, and the
averaging over sizes in the ensemble Eq.~(\ref{trial}) is performed by simply
replacing $\rho \rightarrow \bar \rho$.

Numerical studies have been performed using various forms of the instanton
interaction \cite{Diakonov:1983hh,Diakonov:1995qy}.
The properties of the variational ensemble are not sensitive to the details of the
interaction or other elements of the approximation.
The average instanton size is obtained as $\bar\rho \sim 0.3$ fm, and the instanton packing fraction as
\begin{align}
\kappa \equiv \pi^2 \bar\rho^4 / \bar R^4 \sim 0.1,
\label{packing_fraction}
\end{align}
consistent with the results of lattice simulations. The small value of the instanton packing fraction
(``diluteness'') justifies the approximations made in the functional integration and provides
a small parameter for organizing the calculation of ensemble averages.

An important feature of the instanton ensemble is that all dynamical scales emerge from the
QCD scale in the running coupling, $\Lambda_{\rm QCD}$. No dimensionful parameters are introduced
in the approximations; the instanton interactions are parametrized by dimensionless
constants \cite{Diakonov:1983hh,Diakonov:1995qy}.
As a consequence, all dynamical scales in the low-momentum sector ($k < \bar\rho^{-1}$),
including the average size $\bar\rho$, can be expressed as powers of the instanton density $N/V$.
In this way the instanton density can be regarded as the fundamental scale in low-momentum dynamics,
and all other scales arise from it through the nonlinearity of the
dynamics \cite{Nowak:1989at,Diakonov:1995qy,Kacir:1996qn}.
This fact is essential for the realization of conformal symmetry breaking and the trace anomaly
in the instanton vacuum (see Sec.~\ref{subsec:trace}).

\begin{figure}[t]
\centering{
\includegraphics[width=.9\columnwidth]{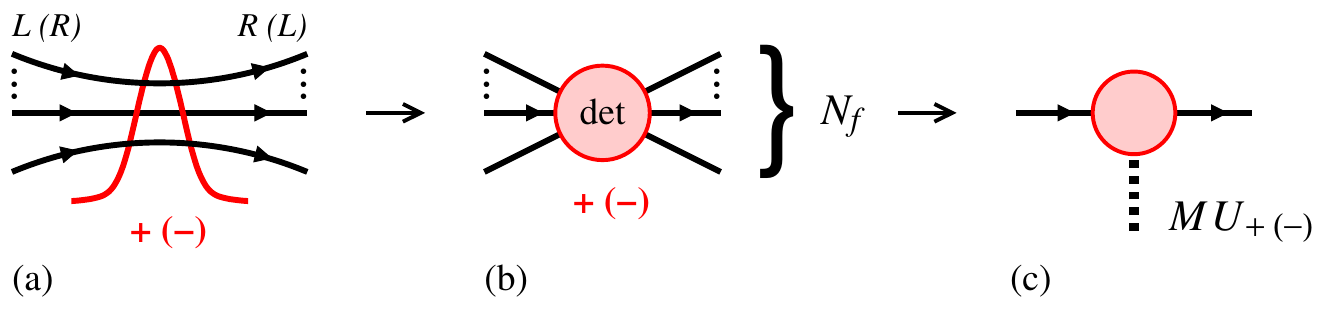}
}
\caption{Fermion interactions induced by instantons.
(a) Fermions in the (anti-) instanton field ($L,R$ denote the chirality).
(b) Flavor interaction from averaging over the color orientation ('tHooft vertex).
(c) Interaction with the chiral field from bosonization.}
\label{fig:instanton_fermion}
\end{figure}

\textit{Fermions and ChSB.}
When fermions are coupled to the instantons, the field of a single (anti-) instanton
induces a zero-virtuality mode of the Dirac operator,
\begin{align}
[i\slashed{\partial}_x + A_\pm (x|z, O, \rho)] \, \Phi_\pm (x|z, O, \rho) = 0.
\end{align}
The zero mode wave function $\Phi_\pm$ is normalizable, localized at the position of the instanton,
and depends on the collective coordinates of the instanton field. The zero mode has definite chirality,
$\gamma_5 \Phi_\pm = \pm \Phi_\pm$. The interaction of the fermion fields
with the zero mode of a single instanton is described by the vertex created by the projector
on the zero mode
\begin{align}
V_\pm (z, O, \rho) [\psi^\dagger, \psi] \equiv
\int d^4 \, x' \psi^\dagger (x') \, i\slashed{\partial} \Phi_\pm (x'|...) 
\int d^4 x \, \Phi_\pm^\dagger (x) \, i\slashed{\partial} \psi (x|...),
\label{zero_mode_projector}
\end{align}
where $\psi^\dagger \equiv i \bar\psi$. In the presence of $N_f \geq 1$ light flavors, the instanton
interacts with all of them (see Fig.~\ref{fig:instanton_fermion}a).
Averaging over the color orientation of the instanton, one obtains a vertex of the form 
\begin{align}
& \int dO \, \prod_{f}^{N_f}
V_\pm (z, O, \rho) [\psi_f^\dagger, \psi_f]
= \textrm{const} \times \textrm{det} \, {\psi^\dagger} (z) \, \overleftarrow{F}
\gamma_{\pm}
\overrightarrow{F} \,
\psi (z),
\label{vertex}
\end{align}
where $\gamma_{\pm} \equiv (1 \pm \gamma_5)/2$ is the chiral projector.
The vertex has the characteristic form of the determinant of the $N_f \times N_f$ matrix
formed by the flavor components of the quark fields and involves all light flavors
in the system  ('tHooft vertex, see Fig.~\ref{fig:instanton_fermion}b) \cite{tHooft:1976snw}.
In Eq.~(\ref{vertex})
\begin{align}
\overrightarrow{F} \psi(z) 
&\equiv \int\!\frac{d^4p}{(2\pi)^4} e^{i p \cdot z} F(p) \psi (p),
\\
\psi^\dagger (z) \overleftarrow{F}
&\equiv \int\!\frac{d^4p'}{(2\pi)^4} e^{-i p' \cdot z} \psi^\dagger (p') F(p'),
\end{align}
where $F(p)$ is a form factor arising from the wave function of the zero mode
in momentum representation, with $F(0) = 1$, and $F(p) \rightarrow 0$ for
$p > \rho^{-1}$ \cite{Diakonov:1995qy}.
The vertex thus has a finite range, given by the instanton size $\rho$.
%
%

In the ensemble with a finite density of instantons, the fermion zero modes cause ChSB.
The phenomenon can be understood as the delocalization of the zero modes
created by the individual instantons and is analogous to band formation in the electron
structure of solids \cite{Diakonov:2002fq,Schafer:1996wv}.
It can be demonstrated by studying the propagation of fermions in the multi-instanton
background using Green function \cite{Diakonov:1985eg,Pobylitsa:1989uq} or effective action
methods \cite{Diakonov:1986aj,Diakonov:1995qy,Kacir:1996qn}.
In the large-$N_c$ limit, the functional integral over the fermions can be computed
in the saddle point approximation. A non-trivial saddle point appears, characterized by a
dynamical quark mass $M$ of parametric order $M^2 \sim \kappa \bar\rho^{-2}$,
with a numerical value $M \sim$ 0.3--0.4 GeV.

The functional integral can be bosonized by introducing meson
fields $\mathcal{M}_\pm (x)$, which are $N_f \times N_f$ matrices in flavor. At the saddle point
\begin{align}
\mathcal{M}_\pm (x)_{f'f} \propto \left\langle \textrm{det}'_{f'f} \;
\psi^\dagger (x) \overleftarrow{F} \gamma_{\pm}
\overrightarrow{F} \psi(x) \right\rangle ,
\label{meson}
\end{align}
where $\langle ... \rangle$ denotes the average over the fermion fields and
$\textrm{det}'_{f'f}$ is the minor of the flavor determinant with row $f$ and column $f'$ removed.
This makes it possible to convert
the multifermion interaction at the saddle point to an interaction of the quarks with the meson field,
\begin{align}
\textrm{det} \; \psi^\dagger (x) \, \overleftarrow{F} \gamma_{\pm}
\overrightarrow{F} \psi(x)
\;\; \rightarrow \;\;
\psi^\dagger (x) \, \overleftarrow{F} \gamma_{\pm} \mathcal{M}_{\pm} (x) \overrightarrow{F} \psi(x).
\end{align}
In many applications the meson field can be restricted to the chiral degrees of freedom and is
parametrized as
\begin{align}
\mathcal{M}_{\pm}(x) = M U_{\pm}(x), \hspace{2em} U_{\pm}(x) = e^{\pm i\pi^a(x) \tau^a},
\end{align}
where $U_\pm$ are $SU(N_f)$ unitary unimodular matrices, $\pi^a$ is the pion field, and $\tau^a$
are the generators of the $SU(N_f)$ algebra.
At the saddle point the effective action of the fermions can then
be represented as \cite{Diakonov:1985eg,Diakonov:1986aj,Diakonov:1995qy,Kacir:1996qn}.
\begin{align}
&S_{\rm eff}(x) = \int d^4 \! x \, \psi^\dagger(x) \left[ -i \slashed{\partial}
- i M \overleftarrow{F} U^{\gamma_5}(x) \overrightarrow{F} \right] \psi (x) ,
\label{S_eff}
\\
&U^{\gamma_5}(x) \equiv \gamma_+ U_+(x) + \gamma_- U_- (x) = e^{i \gamma_5 \pi^a(x) \tau^a}.
\label{U_gamma_5}
\end{align}
It describes the coupling of the massive quarks to the chiral meson field and captures the
low-energy dynamics emerging from ChSB ($k < \bar\rho^{-1}$). The form of the coupling is
dictated by chiral invariance and can be derived from general considerations \cite{Diakonov:1983bny}.
The instanton vacuum provides the dynamical mechanism of ChSB, predicts the value of the
dynamical quark mass, and defines the range of the effective interaction.

Hadronic correlation functions in the effective theory can be computed in the $1/N_c$
expansion. Meson correlation functions exhibit
the quasi massless pion pole in the pseudoscalar-isovector channel, and the massive
$\eta'$ pole in the isoscalar channel [in this channel the pseudoscalar $U_1$ degrees
of freedom in the meson field Eq.~(\ref{meson}) must be retained]
\cite{Diakonov:1985eg,Diakonov:1986aj,Kacir:1996qn}.
Baryon correlation functions are characterized by a classical chiral field (``soliton''),
in which the quarks move in independent-particle orbits \cite{Diakonov:1987ty}.
The calculation of nucleon matrix elements of quark operators (vector/axial currents,
scalars) in this approach has been discussed extensively in the literature;
see Ref.~\cite{Christov:1995vm} for a review.

Instantons convert the QCD color interactions at low energies to effective spin-flavor interactions.
This effect plays an essential role in the emergence of hadron structure from QCD.
The same effect is observed in the gluonic structure of light hadrons induced by instantons
in Sec.~\ref{sec:effop}.

The effective spin-flavor interactions induced by instantons have specific quantum numbers
as encoded in the 'tHooft vertex Eq.~(\ref{vertex}). The interactions occur in the
scalar/pseudoscalar channel and have the characteristic determinantal flavor dependence.
The instanton-induced interactions give rise to ChSB, but this effect is not unique to the
specific quantum numbers and could also be obtained from other effective interactions.
The instanton-induced interactions also give rise to other effects which attest to
the specific spin-flavor quantum numbers, such as the $\eta'$ mass, the differences between
vector and scalar correlation functions, and others \cite{Schafer:1996wv}.
The expression of instantons in low-energy dynamics thus extends beyond ChSB
and can be observed in specific spin-flavor dependent phenomena.

\section{Effective operators from instantons}
\label{sec:effop}
The instanton vacuum allows one to compute correlation functions of QCD operators involving
the gauge fields. Such operators can be converted to ``effective operators'' in the effective
theory of massive quarks with chiral interactions emerging after ChSB \cite{Diakonov:1995qy}.
The effective operators provide a concise representation of the instanton effects in the
QCD operators and enable efficient computation of the hadronic matrix elements.

Consider a gauge-invariant composite QCD operator involving the gauge potential,
$\mathcal{O}[A, \psi^\dagger, \psi]$, normalized at the scale $\mu = \bar\rho^{-1}$.
In the scheme of approximations based on the separation of modes (see Fig.~\ref{fig:instanton_vacuum}a),
the gauge potential in the operator is identified with the classical field of the
superposition of instantons, Eq.~(\ref{gauge_sum}). In leading order of the
packing fraction, the function of the gauge potential can be approximated by
the sum of the functions evaluated in the fields of the individual instantons
\begin{align}
\mathcal{O}[A, \psi^\dagger, \psi] \rightarrow \sum_{I + \bar I} \mathcal{O}[A_I, \psi^\dagger, \psi].
\end{align}
The integration over the collective coordinates of the active instanton, combined with the coupling of
the instanton to the fermions through the the zero mode, converts the gluon operator into
an effective fermion operator. The effective operator is defined such that 
\begin{align}
\left\langle ... \, \mathcal{O}[A, \psi^\dagger, \psi] \, ... \right\rangle_{\textrm{inst}}
\; \stackrel{!}{=} \;
\left\langle ... \, \mathcal{O}_{\rm eff}[\psi^\dagger, \psi] \, ... \right\rangle_{\textrm{eff}} ,
\end{align}
i.e., that the correlation functions of the effective operator in the effective theory
of massive quarks (``after'' integration over instantons) reproduce the correlation functions
the original quark-gluon operator in the instanton ensemble with quarks
(``before'' integration over instantons). The expression of the effective operator is derived
in the saddle-point approximation, going through the same steps as in deriving the effective action.
It is given by \cite{Diakonov:1995qy}
\begin{align}
\mathcal{O}_{\rm eff}[\psi^\dagger, \psi] &= 
\mathcal{N} \; \sum_{\pm} \; \int dz dO d\rho \; d_{\rm eff}(\rho) \; 
\nonumber \\
&\times
\mathcal{O}[A_\pm (z,O,\rho), \psi^\dagger, \psi] \; \prod_{f}^{N_f} V_\pm (z,O,\rho) [\psi_f^\dagger, \psi_f] ,
\label{effop}
\end{align}
where $A_\pm$ is the gauge potential of the single instanton coupling to the operator
and $V_{\pm}$ is its zero mode vertex Eq.~(\ref{zero_mode_projector}). Both depend
on the collective coordinates of the instanton, and the integration connects the QCD operator
with the instanton-induced interactions. The normalization factor $\mathcal{N}$ is determined
within the saddle-point approximation and discussed in Refs.~\cite{Diakonov:1995qy,Balla:1997hf}.

The form of the effective operator depends on the color structure of the QCD operator.
One class of QCD operators have the structure
\begin{align}
\mathcal{O}[A, \psi^\dagger, \psi] = \psi^\dagger (x) \Gamma \psi (x) \mathcal{F}[A],
\label{operator_color_singlet}
\end{align}
where the quark bilinear is a color-singlet and $\mathcal{F}[A]$ is a color-singlet
function of the gauge fields (both $\Gamma$ and $\mathcal{F}$ may carry Lorentz indices,
which are omitted for brevity). In this case the operator in the instanton field does
not depend on the instanton color orientation, and
\begin{align}
\mathcal{F} [A_\pm(z,O,\rho)] = \mathcal{F}_{\pm}(x - z | \rho),
\end{align}
where $\mathcal{F}_{\pm}$ is a scalar function of the coordinates and the size $\rho$.
The color average in Eq.~(\ref{effop}) is then the same as in the fermion vertex Eq.~(\ref{vertex}),
and the gluon part of the operator becomes the 'tHooft vertex,
\begin{align}
\mathcal{O}_{\rm eff}[\psi^\dagger, \psi] &= \psi^\dagger (x) \Gamma \psi (x)
\times \mathcal{N} \; \sum_{\pm} \int d^4 z \; \mathcal{F}_{\pm} (x - z | \bar\rho)
\nonumber \\
&\times \; \textrm{det} \, {\psi^\dagger} (z) \, \overleftarrow{F} \gamma_{\pm}
\overrightarrow{F} \, \psi (z).
\label{effective_color_singlet}
\end{align}
A similar form applies to pure gluon QCD operators without the quark bilinear,
$\mathcal{O}[A] = \mathcal{F}[A]$ \cite{Weiss:2021kpt}.

%
%
\begin{figure}[t]
\centering{
\includegraphics[width=1.\columnwidth]{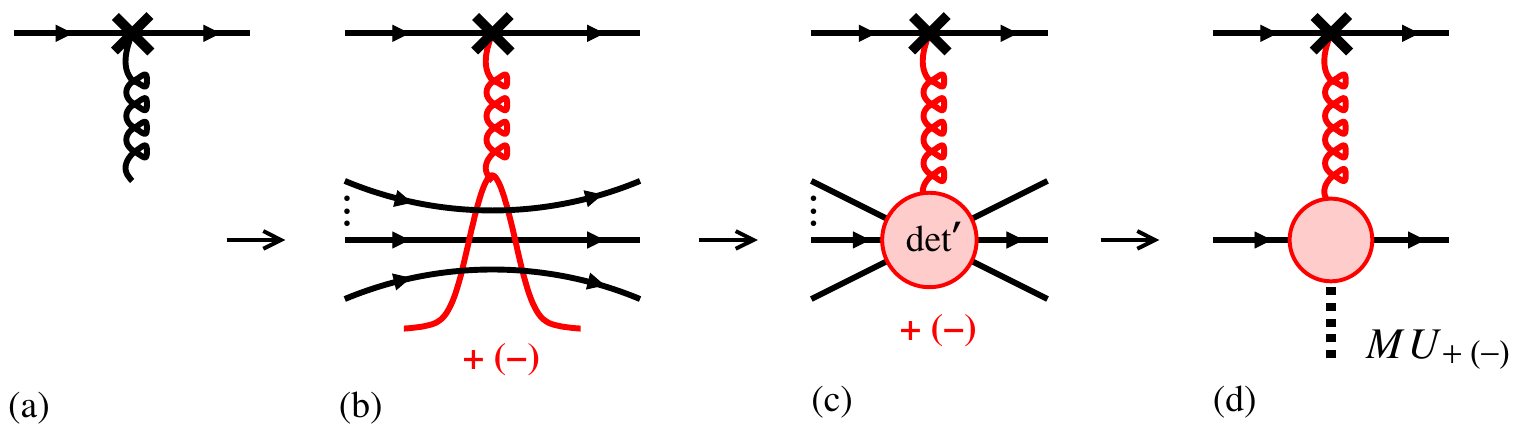}
}
\caption{Effective operators from instantons [here for a color-octet QCD operator
of the form Eq.~(\ref{operator_color_octet})].
(a) QCD quark-gluon operator. (b) Operator with the gluon field evaluated in an (anti-) instanton.
(c) Effective fermion operator from averaging over the collective coordinates.
(d) Bosonized form of the effective operator.}
\label{fig:instanton_effop}
\end{figure}
Another class of QCD operators have the structure
\begin{align}
\mathcal{O}[A, \psi^\dagger, \psi] =
\psi^\dagger (x) \Gamma \frac{\lambda^a}{2} \psi (x) \; \mathcal{F}^a [A](x),
\label{operator_color_octet}
\end{align}
where the quark bilinear is the color-octet current of the quark field and $\mathcal{F}^a [A]$
is a color-octet function of the gauge potential (see Fig.~\ref{fig:instanton_effop}a).
In the field of the instanton the
color-octet function takes the form (see Fig.~\ref{fig:instanton_effop}b)
\begin{align}
\mathcal{F}^a [A_\pm (z,O,\rho)](x) = O^{ab} \eta_{\mp\mu\nu}^b \mathcal{F}_{\pm\mu\nu}(x - z|\rho),
\end{align}
where $\eta_{\mp\mu\nu} \equiv \bar\eta_{\mu\nu}, \eta_{\mu\nu}$ are the 'tHooft symbols and
$\mathcal{F}_{\pm\mu\nu}$ is a tensor-valued function of the coordinates and the size $\rho$.
In this case the function of the instanton field depends on the color orientation,
and the average in Eq.~(\ref{effop}) entangles the function with the zero mode projector.
The effective operator now becomes (see Fig.~\ref{fig:instanton_effop}c)
\begin{align}
& \mathcal{O}_{\rm eff}[\psi^\dagger, \psi] = \psi^\dagger (x) \Gamma \frac{\lambda^a}{2} \psi (x)
\times \mathcal{N} \; \sum_{\pm} \int d^4 z \; \mathcal{F}_{\pm\mu\nu}(x - z | \bar\rho)
\nonumber \\
& \times \sum_{ff'} \; 
{\psi^\dagger}_{\!\! f'} (z) \, \overleftarrow{F} \frac{\lambda^a}{2}
\sigma_{\mu\nu} \gamma_{\pm} \overrightarrow{F} \, \psi_f (z)
\; \textrm{det}'_{f'f} \, {\psi^\dagger} (z) \, \overleftarrow{F} \gamma_{\pm}
\overrightarrow{F} \, \psi (z).
\label{effective_color_octet}
\end{align}
The flavor determinant gets ``differentiated'', and one of the quark bilinears is
now the color-octet Lorentz-tensor projection of the quark fields.

The effective operators induced by instantons are originally obtained as multi-fermion operators,
Eqs.~(\ref{effective_color_singlet}) and (\ref{effective_color_octet}). When used
in correlation functions in the bosonized effective theory Eq.~(\ref{S_eff}),
the multi-fermion effective operators
can be converted to bosonized form, applying the same techniques as in the bosonization of the
effective action, see Eq.~(\ref{meson}). The bosonized form the color-octet effective operator
Eq.~(\ref{effective_color_octet}) is (see Fig.~\ref{fig:instanton_effop}d)
\begin{align}
\mathcal{O}_{\rm eff}[\psi^\dagger, \psi] &= \psi^\dagger (x) \Gamma \frac{\lambda^a}{2} \psi (x)
\; \times \frac{iM}{N_c} \; \sum_{\pm} \; \int d^4 z \; \mathcal{F}_{\pm\mu\nu}(x - z | \bar\rho )
\nonumber \\
& \times \psi^\dagger (z) \, \overleftarrow{F} \frac{\lambda^a}{2} \sigma_{\mu\nu} U_{\pm}(z) \gamma_{\pm}
\overrightarrow{F} \, \psi (z).
\label{effective_color_octet_bosonized}
\end{align}
Here the coefficient is given by dynamical quark mass $M$; its value is unambiguously determined
within the scheme of approximations.
This shows the close connection between the effective dynamics and the effective operators
in the instanton vacuum.

Hadronic matrix elements are computed by inserting the effective operators in correlation functions
and evaluating them using the same methods as for quark operators ($1/N_c$ expansion); see
Refs.~\cite{Balla:1997hf,Dressler:1999zi} for an overview.

The effective operators illustrate the conversion of color interactions to spin-flavor interactions
by the instantons, as already observed in the effective dynamics (see Sec.~\ref{sec:dynamics}).
The different color structure of the QCD operators Eqs.~(\ref{operator_color_singlet}) and
(\ref{operator_color_octet}) gives rise to a different spin-flavor structure of the effective operators
Eqs.~(\ref{effective_color_singlet}) and (\ref{effective_color_octet}). It implies a spin-flavor
dependence of the gluonic structure of hadrons, as can be seen in the hadronic matrix elements.

The expressions Eqs.~(\ref{effop}) and following apply to the effective operators in leading order
of the packing fraction, where the gauge field in the QCD operator is that of a single instanton.
Recent work has derived the effective operators including also instanton-antiinstanton pairs,
which give rise to additional spin-flavor structures (see Sec.~\ref{sec:beyond}) \cite{Liu:2024rdm,Liu:2025ldh}.

\section{Gluonic structure from instantons}
\label{sec:gluonic}
\subsection{Twist-2 parton distributions and momentum sum rule}
The effective operator method can be used to evaluate hadronic matrix elements of various
QCD quark-gluon and pure gluon operators in the instanton vacuum.
It is helpful to organize the discussion according to the twist (= mass dimension minus spin)
of the operators, as this property is important for the size of the instanton effects
and the parametric order of the effective operators. In the following the QCD operators and the instanton-induced
effective operators are presented in Minkowskian form (4-vectors, fields, gamma matrices),
to facilitate comparison with phenomenology; see Ref.~\cite{Schafer:1996wv,Dressler:1999zi} for the
Euclidean-Minkowskian correspondence.

Twist-2 QCD operators describe scaling contributions to the DIS structure functions.
The twist-2 quark and gluon operators of spin-2 are the rank-2
symmetric traceless tensors
\begin{align}
\mathcal{O}_f^{\alpha\beta}(x) &\equiv
\frac{1}{2}
\bar\psi_f (x) \gamma^{\{\alpha} i \nabla^{\beta\}} \, \psi_f (x) - \textrm{trace}
\nonumber \\
&=
\frac{1}{2}
\bar\psi_f (x) \gamma^{\{\alpha} \left( i \partial^{\beta\}}
+ \frac{\lambda^a}{2} (A^{a})^{\beta \}}(x) \right)
\, \psi_f (x) - \textrm{trace},
\label{twist2_quark}
\\
\mathcal{O}_g^{\alpha\beta}(x) &\equiv
F_\gamma^{\; \{\alpha}(x) F^{\beta \} \gamma}(x) - \textrm{trace},
\label{twist2_gluon}
\end{align}
where $\{\alpha\beta\} \equiv \alpha\beta + \beta\alpha$.
The effective operators can be determined using the methods summarized in Sec.~\ref{sec:effop}.
The gluon part of the twist-2 quark operator Eq.~(\ref{twist2_quark}) is of the color-octet form
operator Eq.~(\ref{operator_color_octet}), and the effective operator is given by
the general formula Eq.~(\ref{effective_color_octet}). Explicit calculation shows that,
when the multifermion effective operator is inserted in correlation functions
and the quark fields are contracted, the result is of the order $M^2\bar\rho^2 \sim \kappa$
and thus suppressed in the packing fraction Eq.~(\ref{packing_fraction}) \cite{Balla:1997hf,Kim:2023pll}.
The twist-2 gluon operator Eq.~(\ref{twist2_gluon}) is zero in the field of one instanton,
and its effective operator is zero in leading order of the packing fraction.
Altogether the twist-2 effective operators are obtained as
\begin{alignat}{2}
(\mathcal{O}_f^{\alpha\beta})_{\rm eff} (x) &= \frac{1}{2}
\bar\psi_f(x) \gamma^{\{\alpha} \partial^{\beta \}} \psi_f(x)
- \textrm{trace} && + \mathcal{O}(\kappa),
\label{effective_twist2_quark}
\\
(\mathcal{O}_g^{\alpha\beta})_{\rm eff}(x) &=
\hspace{2em} 0 && + \mathcal{O}(\kappa).
\label{effective_twist2_gluon}
\end{alignat}
The twist-2 quark operators are $\mathcal{O}(1)$ in the instanton packing fraction.
The effect of the gauge potential in the covariant derivative of the QCD operators is suppressed,
and the effective operator is given by the twist-2 operator in the massive quark fields
formed with ordinary derivatives. The twist-2 gluon operator is $\mathcal{O}(\kappa)$
and suppressed in the packing fraction. These conclusions follow from the symmetry
properties of the gauge potential/field of single instanton, in particular its
$O(4)$ rotational covariance \cite{Balla:1997hf,Kim:2023pll}.

The spin-2 twist-2 operators constitute the spin-2 part of the QCD EMT.
Their forward matrix elements (zero momentum transfer) in the nucleon or pion state
define the light-cone momentum fraction carried by quarks/antiquarks and gluons in the hadron,
\begin{align}
\langle p | \, {\textstyle \sum_f} \mathcal{O}_f^{\alpha\beta}(0) \, | p \rangle
&= 2 A_q (p^{\alpha} p^{\beta} - \textrm{trace}),
\\
\langle p | \mathcal{O}_g^{\alpha\beta}(0) \, | p \rangle
&= 2 A_g (p^{\alpha} p^{\beta} - \textrm{trace}).
\end{align}
The effective operators Eqs.~(\ref{effective_twist2_quark}) and
(\ref{effective_twist2_gluon}) imply that
\begin{align}
A_q = 1 + \mathcal{O}(\kappa),
\hspace{2em} A_g = \mathcal{O}(\kappa),
\end{align}
so that the light-cone momentum sum rule is satisfied in leading order of the packing fraction,
\begin{align}
A_q + A_g = 1 + \mathcal{O}(\kappa).
\end{align}
This is a crucial test of the consistency of the approximations. In the instanton vacuum the
momentum sum rule is saturated by quarks and antiquarks in leading order of the packing fraction,
and gluons are suppressed.

These findings can be generalized to the quark and gluon twist-2 operators of spin $n > 2$,
\begin{align}
\mathcal{O}_f^{\alpha_1...\alpha_n} &= \bar\psi_f(x) \gamma^{\{\alpha_1}
\nabla^{\alpha_2}... \nabla^{\alpha_n\}} \psi_f (x) - \textrm{traces},
\\
\mathcal{O}_g^{\alpha_1...\alpha_n} &=
F_{\gamma}^{\;\;\{\alpha_1} D^{\alpha_2}... D^{\alpha_{n-1}} F^{\alpha_n \} \gamma}
- \textrm{traces}.
\end{align}
In leading order of the packing fraction the twist-2 spin-$n$ quark operator is given by the
rank-$n$ symmetric tensor operator in the ordinary derivatives of the massive quark field;
the effect of the gauge potential in the covariant derivatives is suppressed.
The twist-2 spin-$n$ gluon operator is zero in the field of a single instanton.

The nucleon matrix elements of the twist-2 QCD operators define the moments of the nucleon
parton distributions. The effective theory derived from the instanton vacuum can be used to compute
the nucleon parton distributions directly as functions of the light-cone momentum fraction $x$.
Calculations have been performed by evaluating the effective light-cone operators in a nucleon
state at rest \cite{Diakonov:1996sr}, and by using the equivalent formulation in terms of
particle densities in a state with large 3-momentum $p \rightarrow\infty$ \cite{Diakonov:1997vc}.
In this picture the nucleon's partonic structure is carried by the quark and antiquark distributions;
the gluon distribution is suppressed. The antiquark distribution is $\mathcal{O}(1)$
and exhibits a rich spin and flavor dependence, generated by the classical chiral field.
In particular, the picture predicted a large polarized antiquark flavor asymmetry
$\Delta \bar u(x) - \Delta\bar d(x)$, which appears in leading order of the $1/N_c$ expansion
\cite{Diakonov:1996sr,Diakonov:1997vc}.
The prediction agrees with results of experiments in $W^+$ production in polarized $pp$ collisions
at RHIC \cite{STAR:2014afm,STAR:2018fty,PHENIX:2015ade,PHENIX:2018wuz} and a global analysis of
the polarized sea quark distributions \cite{Cocuzza:2022jye}.
It can also be tested with lattice QCD calculations using quasi/pseudo distribution
method \cite{HadStruc:2022nay}.

In summary, at the level of twist-2 structure, and in leading order of the instanton packing fraction,
the instanton fields are ``subsumed'' in the interactions in the effective theory and not
manifest in partonic content. The partonic content is given by quarks and antiquarks.
ChSB determines the effective interactions that create the quark/antiquark distributions
and their spin and flavor dependence. The dynamical mass of the quarks/antiquarks is not manifest
directly in the parton distributions; it is a part of the interactions in the system that define
the parton distributions, not an elementary property of the particles being measured
by the partonic operators. This picture is specific to twist-2 structure and is
qualitatively different in higher-twist structure (see Sec.~\ref{subsec:twist3} and following).

Recent developments enable computation of the twist-2 quark and gluon densities at next-to-leading
order of the packing fraction, including effects of instanton-antiinstanton molecules
(see Sec.~\ref{sec:beyond})
\cite{Liu:2024rdm,Shuryak:2021fsu,Shuryak:2021hng}. Results for the momentum fractions $A_q$ and $A_g$
in the pion show numerically small $\mathcal{O}(\kappa)$ contributions \cite{Liu:2024jno}. Explaining
the nucleon's twist-2 gluon density remains a prime task. Fits to DIS data with valence-like
input densities at low scales $\mu \gtrsim$ 0.5 GeV \cite{Gluck:1998xa}
give large gluon momentum fractions $A_g \sim$ 0.3-0.4, showing the need
for sizable contributions from mechanisms other than single instantons.

\subsection{Twist-3 angular momentum and spin-orbit interactions}
\label{subsec:twist3}
Twist-3 QCD operators appear in the decomposition of the QCD angular momentum in spin and orbital
contributions, and in the description of quark spin-orbit correlations in the nucleon.
The non-forward matrix elements of these operators receive $\mathcal{O}(1)$ contributions
from instantons and represent a unique case of gluonic structure induced by instantons \cite{Kim:2023pll}.

The twist-3 QCD operator with natural parity is given by the antisymmetric rank-2 tensor
\begin{align}
\mathcal{O}^{\alpha\beta}(x) &\equiv
\frac{1}{2}
\bar\psi (x) \gamma^{[\alpha} i \overleftrightarrow{\nabla}^{\beta]} \, \tau \, \psi (x)
\nonumber \\
&=
\frac{1}{2}
\bar\psi (x) \gamma^{[\alpha} \left( i \overleftrightarrow{\partial}^{\beta]}
+ \frac{\lambda^a}{2} (A^{a})^{\beta ]}(x) \right)
\, \tau \, \psi (x),
\label{twist3_natural}
\end{align}
where $\overleftrightarrow{\partial} = \frac{1}{2}(\overrightarrow{\partial} - \overleftarrow{\partial})$
and $[\alpha\beta] \equiv \alpha\beta - \beta\alpha$. $\tau$ denotes a flavor matrix and
can be singlet ($\tau = 1$) or non-singlet ($\tau = \tau^a, a=1,2,3$ for $N_f = 2$).
The operator Eq.~(\ref{twist3_natural}) with the flavor-singlet matrix represents the antisymmetric part
of the QCD EMT [the symmetric part is
given by Eq.~(\ref{twist2_quark})], and its non-forward matrix elements describe the spatial distribution
of quark spin in hadrons \cite{Lorce:2017wkb}. 

%
%
\begin{figure}[t]
\centering{
\includegraphics[width=.95\columnwidth]{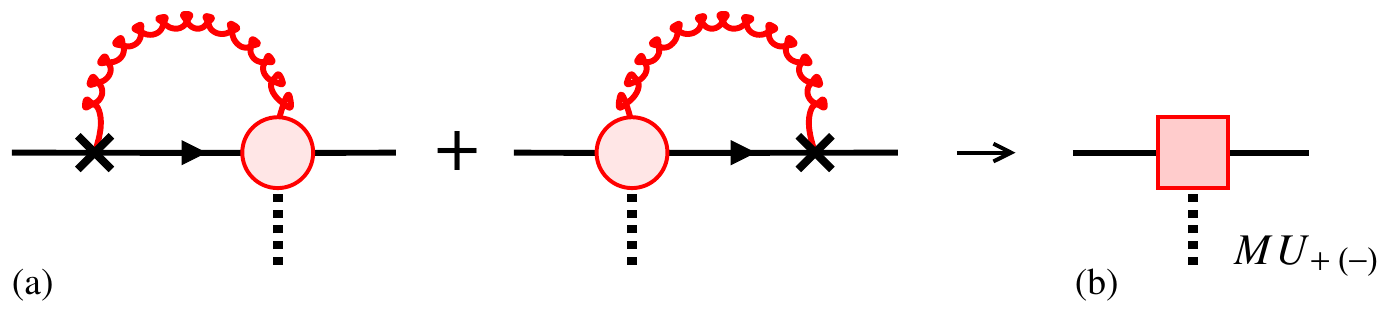}
}
\caption{(a) Contraction of the bosonized effective operator (see Fig.~\ref{fig:instanton_effop}d).
(b) Resulting chiral quark operator. The dashed lines denote the chiral meson field.}
\label{fig:instanton_effop_loop}
\end{figure}
The QCD operator Eq.~(\ref{twist3_natural}) contains the gauge potential in the covariant derivative.
The gauge-potential-dependent term is of the form Eq.~(\ref{operator_color_octet}), and the effective
operator in the instanton vacuum is given by Eq.~(\ref{effective_color_octet_bosonized}), where
the function $\mathcal{F}_{\pm\mu\nu}$ now is the instanton gauge potential \cite{Kim:2023pll}.
Equation~(\ref{effective_color_octet_bosonized}) represents the effective operator as a four-fermion operator,
formed by the product of two color-octet in the background of the chiral field
(see Fig.~\ref{fig:instanton_effop}d).
When inserted in hadronic correlation functions, the fields in the color-octet currents in operator
are contracted, reducing the operator to a two-fermion operator
(see Fig.~\ref{fig:instanton_effop_loop}a). The loop integral resulting from the contraction
can be computed and is parametrically large $\sim \bar\rho^{-2}$ (it would be quadratically
divergent for pointlike vertices and is rendered finite by zero mode form factors).
The integral $\sim \bar\rho^{-2}$ compensates a factor $\sim \bar\rho^{2}$ contained
in the function $\mathcal{F}_{\pm\mu\nu}$ and gives a result that is independent of
the instanton size (see Fig.~\ref{fig:instanton_effop_loop}b).
Altogether the effective operator for the twist-3 QCD operator Eq.~(\ref{twist3_natural})
is obtained as \cite{Kim:2023pll} 
\begin{align}
(\mathcal{O}^{\alpha\beta})_{\rm eff} (x)
&= \frac{1}{2} \bar\psi (x) \left( \gamma^{[\alpha} i \overleftrightarrow{\partial}^{\beta]} \tau
+ \frac{iM}{2} \sigma^{\alpha\beta} [\tau , U^{\gamma_5}(x) ] \right) \psi (x).
\label{twist3_natural_effective}
\end{align}
The first term results from the quark field derivatives in the QCD operator Eq.~(\ref{twist3_natural}).
The second term results from the gauge potential in the QCD operator through the instanton.
The instanton converts the color interaction in the QCD operator to a spin-flavor interaction
of the quark with the chiral field in the effective operator, with a coefficient given by the
dynamical quark mass. (Here the effective operator is presented for energies/momenta
$p \sim M \ll \bar\rho^{-1}$, for which the zero mode form factors in the external quark momenta
can be neglected, $F \rightarrow 1$.)

The twist-3 QCD operator Eq.~(\ref{twist3_natural}) satisfies an operator relation
due to the QCD equations of motion. Using the QCD equations of motion for the quark fields,
\begin{align}
\overrightarrow{\slashed{\nabla}}\psi(x) = 0,
\hspace{2em}
\bar\psi(x)\overleftarrow{\slashed{\nabla}} = 0,
\end{align}
the QCD operator Eq.~(\ref{twist3_natural})
can equivalently be expressed as the total derivative of the QCD axial vector current,
\begin{align}
\mathcal{O}^{\alpha\beta}(x)
&= -\frac{1}{4}
\epsilon^{\alpha\beta\gamma\delta} \partial_\gamma \left[\bar\psi (x) \gamma_\delta \gamma_5 \tau \psi (x) \right],
\label{eom_natural}
\end{align}
where $\partial_\gamma [...]$ denotes the total derivative.
The effective operator Eq.~(\ref{twist3_natural_effective}) satisfies the same relation
in the effective theory of massive quarks with chiral interactions \cite{Kim:2023pll}.
Using the equations of motion of the quark fields in the effective theory (here in Minkowskian convention)
\begin{align}
[ i \overrightarrow{\slashed{\partial}} - M U^{\gamma_5}(x)] \psi (x) = 0,
\hspace{2em}
\bar\psi (x) [ -i \overleftarrow{\slashed{\partial}} - M U^{\gamma_5}(x)]  = 0,
\label{eom_effective}
\end{align}
the effective operator Eq.~(\ref{twist3_natural_effective}) can be converted to the total derivative
of the axial current in the effective theory
\begin{align}
(\mathcal{O}^{\alpha\beta})_{\rm eff} (x)
&= -\frac{1}{4}
\epsilon^{\alpha\beta\gamma\delta} \partial_\gamma \left[\bar\psi (x) \gamma_\delta \gamma_5 \tau \psi (x)
\right]_{\rm eff}.
\label{eom_natural_effective}
\end{align}
This remarkable result is obtained thanks to the chiral interaction term in the effective operator
induced by instantons. It attests to the consistency of the approximations in the derivation of the
effective action and the effective operator (packing fraction expansion, $1/N_c$ expansion).
The QCD equation of motion in the effective operator is realized because the single instanton
is a solution to the Yang-Mills equation. Similar equation-of-motion relations have been
demonstrated for other higher-twist operators in the instanton vacuum \cite{Balla:1997hf}.
An important consequence is that the results of the effective operator calculations
are the same for different QCD operators related by QCD equations of motion
and thus do not depend on the choice of operator basis used for the QCD operator analysis.

The matrix element of the twist-3 QCD operator Eq.~(\ref{twist3_natural}) describes the
spatial distribution of quark spin in the nucleon \cite{Lorce:2017wkb}.
The effective operator Eq.~(\ref{twist3_natural_effective})
reveals an interesting flavor dependence. The instanton-induced interaction term
in Eq.~(\ref{twist3_natural_effective}) is proportional to the flavor commutator $[\tau , U^{\gamma_5}(x) ]$.
In the flavor-singlet operator ($\tau = 1$) the interaction term is zero. The total spin
distribution is thus not affected by the instanton-induced interactions. This is consistent
with the fact that the total spin distribution can be derived from the EMT
in the effective theory, obtained as the conserved current associated with the space-time symmetries.
In the flavor-nonsinglet operator ($\tau = \tau^3$ for $N_f = 2$) the interaction effect is
non-zero. The spin distributions of individual quark flavors are therefore affected by the
instanton-induced interactions.

%
%
\begin{figure}[t]
\centering
\includegraphics[width=.7\columnwidth]{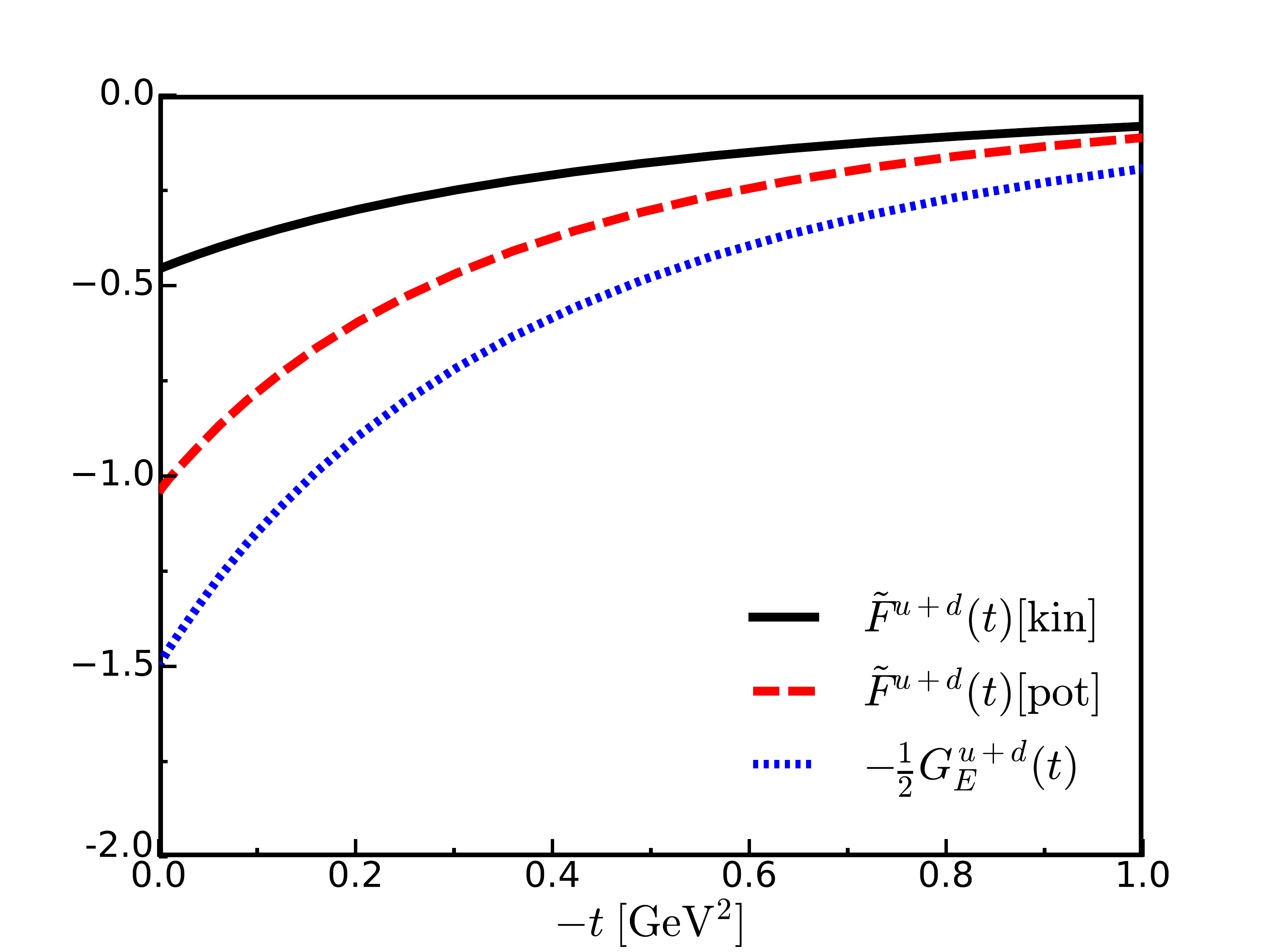}
\caption{Nucleon form factor $\tilde F^{u + d}(t)$ of the twist-3 QCD operator
Eq.~(\ref{twist3_unnatural}) obtained from the instanton vacuum \cite{Kim:2024cbq}.
Black solid line: Contribution of quark field derivative in effective operator
Eq.~(\ref{twist3_unnatural_effective}).
Dashed red line: Contribution of chiral interaction term induced by instantons
in effective operator Eq.~(\ref{twist3_unnatural_effective}).
Dotted blue line: Total result, satisfying the sum rule 
$\tilde F^{u + d}(t) = - \frac{1}{2} G_E^{u + d}(t)$ (nucleon vector form factor).}
\label{fig:spinorbit}
\end{figure}
The twist-3 QCD operator analogous to Eq.~(\ref{twist3_natural}) with unnatural parity,
\begin{align}
\mathcal{O}_5^{\alpha\beta}(x) &\equiv
\frac{1}{2}
\bar\psi (x) \gamma^{[\alpha} \gamma_5 i \overleftrightarrow{\nabla}^{\beta]} \, \tau \, \psi (x),
\label{twist3_unnatural}
\end{align}
describes the spin-orbit correlations of quarks in QCD \cite{Lorce:2014mxa}.
Its effective operator has been be derived, going through same steps as for the natural parity
operator, and is given by \cite{Kim:2023pll}
\begin{align}
(\mathcal{O}_5^{\alpha\beta})_{\rm eff} (x)
&= \frac{1}{2} \bar\psi (x) \left( \gamma^{[\alpha} \gamma_5 i \overleftrightarrow{\partial}^{\beta]} \tau
+ \frac{iM}{2} \sigma^{\alpha\beta} \gamma_5 \{ \tau , U^{\gamma_5}(x) \} \right) \psi (x).
\label{twist3_unnatural_effective}
\end{align}
It exhibits a similar instanton-induced interaction term as the natural-parity operator
Eq.~(\ref{twist3_natural_effective}) and satisfies similar equation-of-motion relations.
In the unnatural-parity operator Eq.~(\ref{twist3_unnatural_effective}) the interaction term
is now proportional to the flavor anticommutator $\{ \tau , U^{\gamma_5}(x) \}$, and is
therefore non-zero also in the flavor-singlet operator. The interaction term plays an essential role
in the theory of spin-orbit correlation of quarks in the nucleon
(see Fig.~\ref{fig:spinorbit}). Its mechanical interpretation in the
mean-field picture of the nucleon in large-$N_c$ limit has been explored in Ref.~\cite{Kim:2024cbq}.

The twist-3 operators Eqs.~(\ref{twist3_natural}) and (\ref{twist3_unnatural}) have hadronic
matrix elements proportional to the momentum transfer $p' - p$ between the states, which
vanish in the forward limit. This can be seen from the fact that the operators can be
converted to total derivative operators using the equations of motion.
Large instanton effects in twist-3 operators have so far only been observed in the
``nonforward'' operators presented here; in other twist-3 operators the effects of the
instanton field are suppressed (see Sec.~\ref{subsec:higher_twist}).

In summary, in twist-3 QCD operators the instanton field can cause $O(1)$ effects, by converting
the color interaction in the QCD operator into a chiral spin-flavor interaction in the effective
operator. This dynamical effect has important consequences for the flavor dependence of the
quark spin distributions and the quark spin-orbit correlations in the nucleon and
should be explored further.

\subsection{Higher-twist quark-gluon correlations and power corrections}
\label{subsec:higher_twist}
Twist-3 and 4 QCD operators appear in power corrections to polarized and unpolarized DIS
structure functions \cite{Shuryak:1981pi,Shuryak:1981kj}.
The instanton effects in these operators depend on the quantum numbers
(spin, isospin) and give rise to a hierarchy of structures, which can be compared
with experimental data.

The twist-3 and 4 operators of dimension 5 with unnatural parity are
\begin{alignat}{2}
\mathcal{O}^{\alpha\beta\gamma}(x)
&= \bar\psi (x) \gamma^{\{\alpha } \widetilde{F}^{\beta\}\gamma}(x) \, \tau \psi(x) - \textrm{traces}
\hspace{2em}
&& \textrm{(twist-3)},
\label{twist3}
\\
\mathcal{O}^{\beta}(x)
&= \bar\psi(x) \gamma_{\alpha} \widetilde{F}^{\beta\alpha}(x) \, \tau \psi (x)
&& \textrm{(twist-4)},
\label{twist4}
\end{alignat}
where $\widetilde{F}^{\beta\gamma} = \frac{1}{2} \epsilon^{\beta\gamma\delta\epsilon} F_{\delta\epsilon}$
is the dual field strength and $\tau$ is a
flavor matrix. The nucleon matrix elements are parametrized as\footnote{The twist-4 matrix element
$f_2$ is defined here with mass dimension $\textrm{(mass)}^2$,
which is natural for the present discussion. In the literature the mass dimension of $f_2$ is usually
absorbed by factor nucleon mass, $f_2(\textrm{our}) = m_N^2 f_2(\textrm{lit})$.}
\begin{align}
\langle p s| \mathcal{O}^{\alpha\beta\gamma}(0) | p s \rangle
&= 2 d_2 ( 2 p^{\{\alpha} p^{\beta\}} s^\gamma - 2 p^{\{ \alpha }
s^{\beta \} } p^{\gamma} - \textrm{traces} ),
\\
\langle p s| \mathcal{O}^{\beta}(0) | p s \rangle &= 2 f_2 s^\beta,
\end{align}
where $s$ is the polarization 4-vector, $s^\mu p_\mu = 0, s^2 = -1$. They describe quark-gluon
correlations in various spin projections in the polarized nucleon
and can be interpreted as electric and magnetic color polarizabilities;
see Ref.~\cite{Kuhn:2008sy} for a review.
$d_2$ and $f_2$ appear in the $1/Q^2$ power corrections to the lowest moment of the
spin structure function $g_1$. $d_2$ also appears in the $x^2$ moment of $g_2$,
at the same level in $1/Q^2$ as the twist-2 matrix element; the contribution of $g_2$
to the DIS cross section is overall power-suppressed by $1/Q^2$ \cite{Ehrnsperger:1993hh}.

The instanton vacuum makes definite predictions for the spin-dependent
higher-twist matrix elements \cite{Balla:1997hf}.
The twist-3 effective operator obtained from Eq.~(\ref{twist3})
has hadronic matrix elements of order $M^2\bar\rho^2 \sim \kappa$,
while the twist-4 effective operator obtained from Eq.~(\ref{twist4}) has
matrix element of order unity,
\begin{align}
d_2 &= \mathcal{O}(\kappa),
\hspace{2em}
f_2 \sim \bar\rho^{-2} = \mathcal{O}(1).
\end{align}
The reason for the different behavior is the $O(4)$ rotational symmetry of the instanton field.
It governs the parametric order of the loop integral appearing in the contraction of the
effective operator (see Fig.~\ref{fig:instanton_effop_loop}a) and causes a qualitative
difference between the results for the twist-3 (= spin-2) and twist-4 (= spin-1) operators.

Numerical predictions for the higher-twist matrix elements from the instanton vacuum
have been obtained and can be compared with experimental extractions.
The small value $d_2 \sim \textrm{few} \times 10^{-3}$ predicted in Ref.~\cite{Balla:1997hf}
is consistent with the results of $g_2$ measurements at SLAC and JLab; see Ref.~\cite{Sato:2016tuz}
for a global analysis. They are also consistent with modern lattice QCD calculations using
nonperturbative renormalization \cite{Gockeler:2005vw}. The value $f_2^{u - d} \sim 0.2$ GeV$^2$
predicted in \cite{Balla:1997hf} is consistent with the extraction of higher-twist corrections
to $g_1$ in Ref.~\cite{Sidorov:2006vu}; for estimates of $f_2^{u + d}$ see Ref.~\cite{Lee:2001ug}.
While large uncertainties remain, especially in the extraction
of $f_2$ from power corrections, the pattern appears consistent with the instanton predictions.

Other twist-4 operators appear in the power corrections to the unpolarized structure
functions. The $1/Q^2$ corrections to $F_2$ and $F_L$ involve quark-gluon and 4-quark
operators \cite{Shuryak:1981kj}.
The instanton vacuum predicts that one of the quark-gluon operators has a large matrix element
$\bar\rho^{-2} = \mathcal{O}(1)$,
while the matrix elements of the 4-quark operators are suppressed \cite{Dressler:1999zi}.
As a consequence, large $1/Q^2$ corrections are predicted in $F_L$, and much smaller corrections in $F_2$.
The pattern is consistent with the results of a phenomenological analysis of higher-twist
corrections \cite{Dressler:1999zi}. The twist-4 operators appearing in the neutrino structure
functions $F_{1\nu}$ and $F_{3\nu}$ have also been computed and have matrix elements
$\mathcal{O}(1)$ \cite{Balla:1997hf,Weiss:2002qq}.

In summary, the instanton vacuum allows one to evaluate a range of higher-twist operators
governing power corrections to deep-inelastic processes. It predicts a hierarchy
of higher-twist matrix elements, with ``selection rules'' dictated by the $O(4)$ rotational
symmetry of the instanton field. Power corrections thus show the footprint of the
topological fields in deep-inelastic processes. The approach has been applied to compute
matrix elements of chiral-odd higher-twist operators \cite{Dressler:1999hc}
and non-forward matrix elements of higher-twist operators in generalized parton
distributions \cite{Kiptily:2002nx}. It can also be extended to compute power corrections
in the extraction of quasi-parton distributions \cite{Liu:2021evw}.

\subsection{Trace anomaly and hadron mass decomposition}
\label{subsec:trace}
The twist-4 scalar gluon operator $F^{\mu\nu}F_{\mu\nu}$ appears in the trace
anomaly of the EMT and plays an important role in the hadron mass decomposition in QCD.
The instanton vacuum encodes the trace anomaly in the density fluctuations of instantons and allows
one to compute and interpret its hadronic matrix elements.

At the classical level, QCD is scale invariant up to effects proportional to the light quark masses,
and the trace of the EMT is zero, $T^\mu_{\;\;\mu} = 0 + O(m)$. At the quantum level, scale invariance is
broken by quantum fluctuations, which require an UV cutoff whose presence is felt even after sending
it to infinity (anomaly). The trace of the EMT becomes
\begin{align}
&T^\mu_{\;\;\mu} (x) = \frac{\beta (g)}{4g^4} F^{\mu\nu} F_{\mu\nu} (x)
+ m [1 + \gamma_m(g)] \sum_f \bar\psi_f \psi_f (x) ,
\label{trace_anomaly}
\\
&\frac{\beta (g)}{4g^4} = -\frac{b}{32\pi^2} + \mathcal{O}(g^2),
\hspace{2em} \gamma_m (g) = \mathcal{O}(g^2),
\end{align}
where the coefficient $b$ of the beta function is given in Eq.~(\ref{d_0}), $m$ is the
light quark mass (assumed to be the same for all light flavors here),
and $\gamma_m$ its anomalous dimension. 
This operator relation equates the trace of the EMT with the dimension-4 gluon operator.
It connects the breaking of scale invariance in QCD with the scalar gluon content of hadrons
and is fundamental for hadron structure.
The expectation value of the EMT in the nucleon state at rest,
averaged over the spin, is given by\footnote{The nucleon state is denoted here
by $|n\rangle$, to avoid confusion with the instanton number $N$.}
\begin{align}
\langle n | T^\mu_{\;\;\mu} (0) | n \rangle
= 2 m_n^2 .
\label{trace_nucleon}
\end{align}
This follows from the general form of the nucleon matrix element of the EMT (relativistic covariance, current
conservation $\partial_\mu T^{\mu}_{\;\;\nu} = 0$) and the fact that $T^{00}$ measures the total energy
of the state. Equations~(\ref{trace_anomaly}) and (\ref{trace_nucleon}) imply that
\begin{align}
\frac{\langle n | F^{\mu\nu} F_{\mu\nu}(0) | n \rangle}{2 m_n^2}
= -\frac{32\pi^2}{b} + \mathcal{O}(m).
\label{F2_nucleon}
\end{align}
This is a remarkable statement: The average of the scalar gluon density in the nucleon state,
a quantity arising from nonperturbative dynamics, is constrained by the coefficient of the QCD beta function,
a property of perturbation theory. It calls for a mechanical explanation how this connection is realized.

%
%
\begin{figure}[t]
\begin{tabular}{ll}
\hspace{.1\columnwidth}
\includegraphics[width=.35\columnwidth]{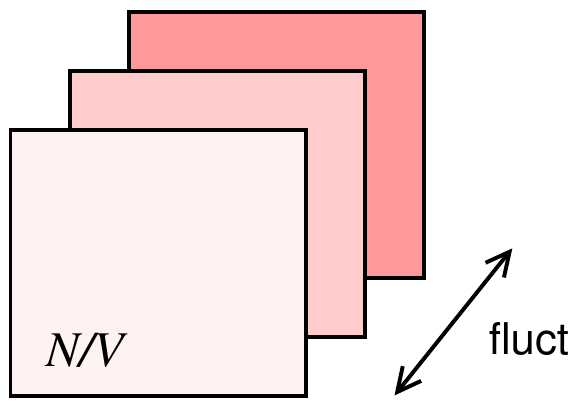}
\hspace{.1\columnwidth}
&
\includegraphics[width=.3\columnwidth]{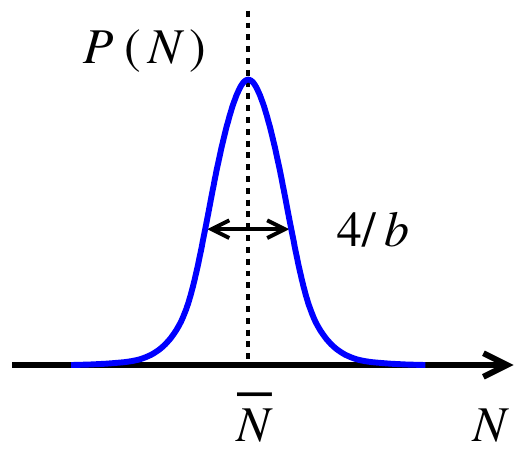}
\\[-1ex]
{\small (a)} & {\small (b)}
\end{tabular}
\caption{(a)~Instanton density fluctuations in the grand canonical ensemble.
(b)~Instanton number distribution Eq.~(\ref{numberdist}).}
\label{fig:instanton_number}
\end{figure}
In the instanton vacuum the trace anomaly is expressed in the fluctuations of the instanton density.
The variational approximation to the QCD vacuum is performed with a grand canonical ensemble
of instantons, with a variable instanton number $N = N_+ + N_-$ fluctuating with a
distribution $P(N)$ (see Fig.~\ref{fig:instanton_number}) \cite{Nowak:1989at,Diakonov:1995qy}.
The ensemble average is now understood as
\begin{align}
\sum_N P(N) \;
\left\langle ... \phantom{0^0_0} \hspace{-1em}  \right\rangle_N ,
\end{align}
where the subscript $N$ denotes the canonical ensemble average used in the derivation
of ChSB (see Sec.~\ref{sec:dynamics}).
The instanton number is proportional to the volume-integrated Euclidean
operator $F^2 = F_{\mu\nu}F_{\mu\nu}$,
\begin{align}
N = \frac{1}{32 \pi^2} \int d^4 x \; F^2 (x),
\end{align}
where $32\pi^2$ is the action of a single instanton.
The instanton number distribution can be inferred from the low-energy theorems for the
vacuum correlation functions of the scalar gluon operator in gluodynamics,
\begin{align}
\left\langle \frac{1}{32 \pi^2}\int F^2 \; \frac{1}{32 \pi^2}\int F^2 \right\rangle
- \left\langle \frac{1}{32 \pi^2}\int F^2 \right\rangle^2
= \frac{4}{b} \left\langle \frac{1}{32 \pi^2}\int F^2 \right\rangle
\end{align}
etc., which are derived by differentiating the renormalized partition function with respect to the
inverse coupling constant \cite{Novikov:1981xi}. It is obtained as \cite{Diakonov:1995qy}
\begin{align}
P(N) \propto \left(\frac{N}{\bar N}\right)^{-bN/4} e^{bN/4} .
\label{numberdist}
\end{align}
The variance of the instanton number fluctuations is controlled by the coefficient of the
beta function,
\begin{align}
\frac{\overline{(N - \bar N)^2}}{\bar N} = \frac{4}{b},
\label{width}
\end{align}
and is known as the topological vacuum compressibility. The instanton number
fluctuations are stronger than Poissonian, attesting to the presence of interactions in the system.

Hadronic matrix elements of $F^2$ are extracted from the 3-point correlation functions
of the gluon operator with interpolating operators for the hadronic states.
In the connected part of the 3-point correlation function the contributions proportional to
the average instanton number $\bar N$ cancel, and the result arises entirely from
the fluctuations of $N$. Schematically,
\begin{align}
&\frac{\langle n| F^2(0) | n \rangle}{\langle n | n \rangle}
= \lim_{T \rightarrow \infty} \frac{\langle J_n (T) F^2(0) J_n (-T)\rangle_{\rm conn}}
{\langle J_n (T) J_n (-T)\rangle}
\nonumber \\
&= \lim_{T \rightarrow \infty} \frac{32\pi^2}{2 T V_3} \;
\frac{\overline{(N - \bar N)^2}}{\bar N} \;
\left.
\frac{ \displaystyle N \frac{d}{dN} \langle J_n (T) J_n (-T)\rangle_N}{\langle J_n (T) J_n (-T)\rangle_N}
\right|_{N = \bar N} ,
\label{connected}
\end{align}
where $J_n$ is the nucleon interpolating operator,
$T$ is the Euclidean time separation, $V_3$ is the spatial volume,
and $\langle n | n \rangle = 2 m_n V_3$
from the normalization of the states \cite{Diakonov:1995qy}.
At large times the 2-point correlation function decays exponentially, with the range
given by the nucleon mass. The logarithmic derivative of the correlation function with respect to the
instanton number becomes the derivative of nucleon mass,
\begin{align}
\langle J_n (T) J_n (-T)\rangle_N \propto e^{-2 m_n T},
\hspace{2em}
\frac{1}{2T}
\frac{ \displaystyle N \frac{d}{dN} \langle ... \rangle_N}{\langle ... \rangle_N}
= - \displaystyle N \frac{dm_n}{dN} .
\end{align}
In the instanton vacuum at zero light quark masses, the only dynamical scale is the instanton density $N/V$,
and hadronic scales arise as powers of this scale according to their naive mass dimension
(see Sec.~\ref{sec:dynamics}). The nucleon mass thus depends on the instanton density as
\begin{align}
m_n = \textrm{[dimensionless constant]} \, \times \, \left( \frac{N}{V} \right)^{1/4}.
\label{nucleon_mass}
\end{align}
Altogether Eqs.~(\ref{connected})--(\ref{nucleon_mass}) determine the nucleon matrix element as
\begin{align}
\frac{\langle n | F^2(0) | n \rangle}{2 m_n^2}
= -\frac{32\pi^2}{b} ,
\label{F2_nucleon_instanton}
\end{align}
in agreement with the general result from the trace anomaly Eq.~(\ref{F2_nucleon}).
This remarkable result comes about because in the instanton vacuum the information
on the beta function is encoded in the instanton number fluctuations Eq.~(\ref{width}),
and the nucleon mass arises as a power of the instanton density.

The result of Eq.~(\ref{F2_nucleon_instanton}) shows several interesting features.
(i)~The expectation value of $F^2$ in the nucleon state is negative, even though
$F^2 > 0$ in the Euclidean metric. This is explained by the fact that the nucleon matrix element
measures the change in the vacuum expectation value of $F^2$ caused by presence of the nucleon,
and the change is negative. (ii)~In the large-$N_c$ limit $b \sim N_c$, see Eq.~(\ref{d_0}).
The matrix element Eq.~(\ref{F2_nucleon_instanton}) is therefore suppressed in $1/N_c$ compared
to its natural size. This circumstance allows for the mixing of gluon and scalar
quark/antiquark modes in the $t$-channel spectral representation of the matrix element
and plays an important role in the generalization to non-zero momentum transfer (see Sec.~\ref{subsec:scalar}).

The trace anomaly Eq.~(\ref{trace_anomaly}) can also be evaluated in the pion state.
In this case the quark mass term cannot be neglected, because the pion mass depends
on the quark mass as $M_\pi^2 \propto m$. One obtains \cite{Liu:2024jno}
\begin{align}
-\frac{b}{32 \pi^2} \frac{\langle \pi | F^2 | \pi \rangle}{2 M_\pi} +
\frac{\langle \pi | \, m \sum_f \bar\psi_f \psi_f | \pi \rangle}{2 M_\pi} = M_\pi ,
\end{align}
where the contribution of $\gamma_m = \mathcal{O}(g^2)$ in the quark mass term
is neglected. It shows that the pion mass arises both from the scalar density of the gluon field
(trace anomaly) and the quark mass times the scalar quark density (sigma term).
The instanton vacuum allows one to compute both terms separately, using the
methods described above, and gives \cite{Liu:2024jno}
\begin{align}
-\frac{b}{32 \pi^2} \frac{\langle \pi | F^2 | \pi \rangle}{2 M_\pi}
&=
\frac{M_\pi}{2} \left[ 1 + \mathcal{O}(m) \right],
\label{pion_trace}
\\
\frac{\langle \pi | \, m \sum_f \bar\psi_f \psi_f | \pi \rangle}{2 M_\pi}
&=
\frac{M_\pi}{2} \left[ 1 + \mathcal{O}(m) \right].
\label{pion_sigma}
\end{align}
It shows that the pion mass arises in equal parts from the trace anomaly and the sigma term.
Note that the pion matrix element of $F^2$, Eq.(\ref{pion_trace}),
vanishes in the chiral limit, highlighting the interplay of conformal and chiral symmetry breaking.
The result for the pion sigma term, Eq.~(\ref{pion_sigma}), agrees with the result of chiral reduction
(soft-pion theorem). Altogether, these results attest to the consistency of the approximations
in the description of conformal and chiral symmetry breaking in the instanton vacuum.

In summary, the instanton vacuum describes the hadronic matrix elements of the trace anomaly
in accordance with the low-energy theorems of conformal and chiral symmetry breaking.
It provides a mechanical picture for the intrusion of the beta function into nonperturbative
hadron structure. As such it represents an essential tool for the study of the hadron mass decomposition
and the interplay of conformal and chiral symmetry breaking in QCD.

\subsection{Scalar gluon form factors and mechanical properties}
\label{subsec:scalar}
Much more information is contained in the form factor of the scalar gluon operator
$F^{\mu\nu} F_{\mu\nu}$ at nonzero momentum transfer.
The form factors of the EMT define the so-called mechanical properties of hadrons,
which have become field of study in their own right; see Refs.~\cite{Lorce:2018egm,Polyakov:2018zvc}
for reviews. The scalar gluon form factor of the nucleon is also measured in heavy quarkonium
photo- and electroproduction near threshold; see Ref.~\cite{Duran:2022xag} and references therein.

The scalar gluon form factors of light hadrons can be computed in the instanton vacuum.
An important effect at nonzero momentum transfer is the mixing of the instanton density
fluctuations with the scalar quark-antiquark modes in the effective dynamics
arising from ChSB (``glueballs'' and ``mesons'' in colloquial terms).
This mixing can be studied within the $1/N_c$ expansion \cite{Kacir:1996qn,Liu:2024rdm,Liu:2024jno}.
It is made possible by the fact that the width of the instanton density fluctuations is
$b \sim 1/N_c$, so that the action of these modes is of the same order in
$1/N_c$ as that of scalar quark-antiquark modes. Practical methods for the computation
of nonforward matrix elements of scalar operators have been developed in the regimes
of momentum transfers of the order $Q \sim M$ (soft regime), where the instantons act collectively,
and $Q \sim \bar\rho^{-1}$ (semi-hard regime), where the scattering process is mediated by single
instantons or pairs \cite{Liu:2024rdm,Liu:2024vkj}.

\begin{figure}[t]
\centering
\includegraphics[width=.9\columnwidth]{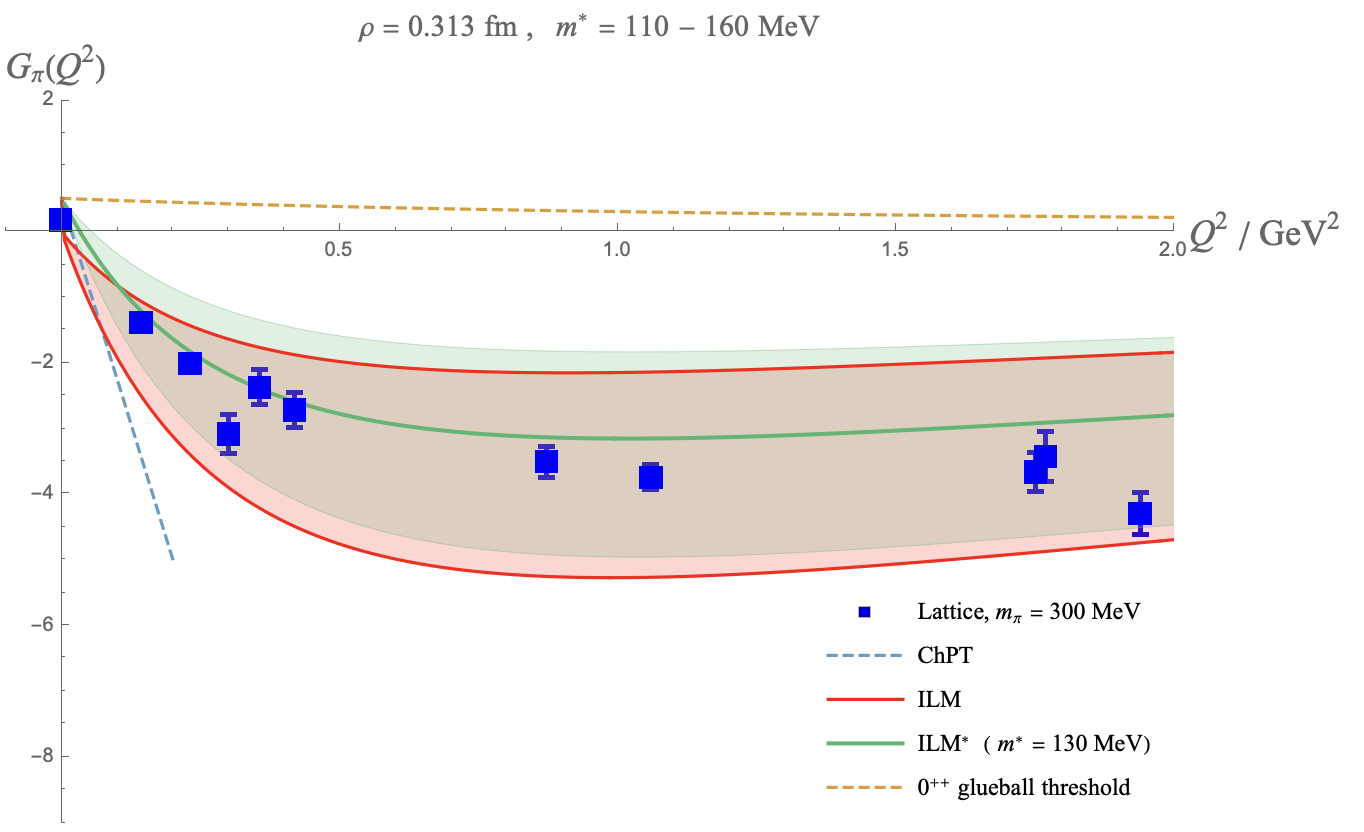}
\caption{Pion scalar gluon form factor from the instanton vacuum \cite{Liu:2024jno}.
Green band: Full instanton vacuum result with uncertainty estimate (see reference for details).
Red bands: Semi-hard contribution $Q \sim 1/\bar\rho$ in instanton vacuum.
Blue points: Lattice QCD results \cite{Wang:2024lrm}.
Blue dashed line: Form factor slope from LO chiral perturbation theory.
Yellow dashed line: $0^{++}$ glueball exchange.}
\label{fig:pion_scalar}
\end{figure}

Fig.~\ref{fig:pion_scalar} shows the results of an instanton vacuum calculation of the
gluon scalar form factor of the pion, defined as [see Eq.~(\ref{trace_anomaly})] 
\begin{align}
G_\pi (Q^2) = \frac{\langle \pi (p') | T^\mu_{\;\;\mu}(0)_{\rm glue} | \pi (p) \rangle}{2 M_\pi^2}
= -\frac{b}{32 \pi^2} \frac{\langle \pi (p') | F^{\mu\nu}F_{\mu\nu} (0) | \pi (p) \rangle}{2 M_\pi^2},
\end{align}
where $Q^2 = -(p' - p)^2$ \cite{Liu:2024jno}. This calculation uses the light-front formulation developed
in Refs.~\cite{Shuryak:2021fsu,Shuryak:2021hng} and subsequent works and includes
instanton pair contributions; see Ref.~\cite{Liu:2024jno} for details.
At zero momentum transfer it reproduces the value $G_\pi (0) = 1/2$
obtained from the instanton density fluctuations (see Sec.~\ref{subsec:trace}).
The instanton vacuum result for the pion scalar gluon form factor agrees well with
recent lattice QCD calculations \cite{Wang:2024lrm,Hackett:2023nkr}.
%
%

In summary, the instanton vacuum predicts the scalar gluon form factors of light hadrons
up to momentum transfers $Q \sim$ few GeV$^2$ based on the nonperturbative gauge field dynamics
abstracted from lattice QCD calculations. It represents an essential tool in the study
of hadron mechanical properties and high-momentum transfer processes.
Future studies could (i)~compute the nucleon and other baryon form factors in the chiral soliton picture
in the large-$N_c$ limit, (ii)~revisit the quark and gluon decomposition of the $\bar c$ form
factor of Ref.~\cite{Polyakov:2018exb} and extend it to nonzero momentum transfer,
(iii)~compute the amplitude of heavy quarkonium production on light hadrons
and other gluonic processes mediated by the instantons.

\subsection{Axial anomaly and pseudoscalar gluon form factors}
The dimension-4 pseudoscalar gluon operator $F^{\mu\nu}\tilde{F}_{\mu\nu}$ represents the
topological charge density of the gauge fields in QCD. The $U(1)_A$ axial anomaly
equates this operator with the divergence of the flavor-singlet axial current,
providing a direct connection between the gauge field topology and the spin-flavor
dynamics of light quarks. The instanton vacuum encodes the axial anomaly through
topological fluctuations of the instanton number, making it possible to compute
and interpret the hadronic matrix elements of $F^{\mu\nu}\tilde{F}_{\mu\nu}$.
The pseudoscalar meson matrix elements ($\eta'$ mass etc.) have been extensively discussed
in the literature \cite{Schafer:1996wv,Diakonov:1995qy}; the present review focuses on the
nucleon matrix element.

In classical QCD the axial currents are conserved in the chiral limit.
In the quantum theory the flavor-singlet axial current acquires an anomalous
divergence due to quantum fluctuations and renormalization,
\begin{align}
{\textstyle \sum_f} \partial^\mu [\bar\psi_f \gamma_\mu\gamma_5 \psi_f](x)
&= \frac{N_f}{16\pi^2} \, F^{\mu\nu} \tilde F_{\mu\nu}(x) +
2 \, {\textstyle \sum_f} m_f \bar \psi_f i \gamma_5 \psi_f(x) .
\label{axial_anomaly}
\end{align}
This operator relation connects the hadronic matrix elements of $F^{\mu\nu}\tilde{F}_{\mu\nu}$
with those of the axial current. The nucleon matrix element of the pseudoscalar
gluon operator is parametrized as
\begin{align}
\langle p' \sigma' | \, F^{\mu\nu} \tilde{F}_{\mu\nu}(0) \, | p \sigma \rangle
&= A_{P}(q^2) \, m_n \bar u' i \gamma_5 u ,
\label{FFdual_matrix}
\end{align}
where $u \equiv u(p, \sigma), u' \equiv u(p', \sigma')$ are the nucleon 4-spinors and $A_P$
is an invariant form factor. Using the axial anomaly Eq.~(\ref{axial_anomaly}) and the
standard representation of the nucleon matrix element of the axial current in terms
of the axial and pseudoscalar form factors, and taking the chiral limit, one obtains
\begin{align}
A_P(0) \; &= \;
2 g_A^{(0)} / N_f ,
\label{FFdual_from_axial}
\end{align}
where $g_A^{(0)}$ is the nucleon flavor-singlet axial coupling. This remarkable relation
connects the nucleon matrix elements of the gluon operator and the light-quark operator
of the axial current.

In the instanton vacuum the axial anomaly is expressed in the fluctuations of the topological
charge of the instanton ensemble, i.e., fluctuations of the difference of the number of
instantons and antiinstantons,
\begin{align}
\Delta \equiv N_+ - N_-.
\end{align}
They can be implemented in the variational approximation with the grand canonical ensemble,
in a similar way as the fluctuations of the total instanton number
$N = N_+ + N_-$ (see Sec.~\ref{subsec:trace}) \cite{Nowak:1989at,Diakonov:1995qy,Kacir:1996qn}.
The fluctuations of $\Delta$ are described by a distribution $P(\Delta)$.
The instanton number difference is equal to the topological charge in the form of
the volume-integrated Euclidean operator $F\tilde F \equiv F_{\mu\nu} \tilde{F}_{\mu\nu}$,
\begin{align}
\Delta = \frac{1}{32 \pi^2} \int d^4 x \; F\tilde{F} (x).
\label{Delta_FFdual}
\end{align}
The distribution $P(\Delta)$ can be inferred from the vacuum correlation function
of the topological charge,
\begin{align}
\left\langle \frac{1}{32 \pi^2}\int F\tilde{F} \; \frac{1}{32 \pi^2}\int F\tilde{F} \right\rangle ,
\end{align}
the so-called topological susceptibility of the QCD vacuum. The topological susceptibility
has been analyzed both in pure gluodynamics and including fermions; it is qualitatively
affected by the presence of light fermions and vanishes in the chiral limit \cite{Schafer:1996wv}.
The distribution $P(\Delta)$ is obtained as \cite{Diakonov:1995qy}
\begin{align}
P(\Delta ) &\propto \exp \left( - \frac{\Delta^2}{2 N} \right) \times
\exp \left( \frac{\Delta^2}{2 V \langle \bar\psi \psi \rangle}
\sum_f^{N_f}  m_f^{-1} \right) .
\label{P_Delta}
\end{align}
The first factor results from gluodynamics (so-called quenched topological susceptibility);
the second factor results from the fermion determinant (note that $\langle \bar\psi \psi \rangle < 0$).
The width of the distribution vanishes when $m_f \rightarrow 0$ for at least one flavor $f$.
For quark masses close to the chiral limit, the width of the second factor
is much smaller than that of the first, and the width of the overall distribution is dominated
by the fermions and their chiral behavior.

The distribution Eq.(\ref{P_Delta}) can also be derived directly from the fermion determinant
in the instanton ensemble with fermions \cite{Diakonov:1995qy}. The instanton ensemble provides
a simple explanation of the vanishing of the topological susceptibility in the chiral limit.
In instanton background fields with nonzero topological charge, $\Delta \neq 0$
and $N_+ \neq N_-$, the fermion spectrum after ChSB
contains ``unbalanced'' zero modes, which cause the determinant to vanish in the chiral limit
(the functional determinant is the product of the eigenvalues of the modes).

The hadronic matrix elements of $F\tilde{F}$ can be extracted from 3-point correlation functions,
using similar methods as for $FF$ (see Sec.~\ref{subsec:trace}).
The average is performed with the grand canonical ensemble.
The $\Delta$ fluctuations connect the gluon operator Eq.~(\ref{Delta_FFdual}) with the
effective dynamics of the light quarks, in a way that the $U(1)_A$ anomaly is realized;
see Ref.~\cite{Diakonov:1995qy} for details. The chiral singularity of the $\Delta$ fluctuation
width is canceled, and a stable result is obtained in the chiral limit.
The nucleon matrix element Eq.~(\ref{FFdual_matrix}) is obtained consistently
with the $U(1)_A$ anomaly Eq.~(\ref{FFdual_from_axial}), with $g_A^{(0)}$ given by
the nucleon axial coupling predicted by the effective dynamics of light quarks.
The consistent realization of this subtle relation between gluon and light quark dynamics
is a major accomplishment for an effective description of low-energy QCD.
\footnote{The nucleon matrix element Eq.~(\ref{FFdual_matrix}) vanishes for zero momentum
transfer $q \rightarrow 0$ and has to be computed using the position-dependent local operator
$F\tilde{F}(x)$. A careful procedure is needed when integrating the position over a
finite volume and associating the operator with the topological charge of the ensemble,
requiring considerations beyond the treatment of Ref.~\cite{Diakonov:1995qy}; see
Ref.~\cite{Weiss:2021kpt} for a discussion.}

The form factors of $F\tilde{F}$ and similar operators at finite momentum transfer
can be computed using the methods developed in Refs.~\cite{Kacir:1996qn,Liu:2024rdm,Liu:2024jno}.
The pseudoscalar form factors at momentum transfers $Q \sim M$ are governed by the mixing of gluon
and pseudoscalar quark-antiquark modes, similar to the scalar gluon form factors
(see Sec.~\ref{subsec:scalar}).

In summary, the instanton vacuum describes the hadronic matrix elements of $F\tilde{F}$
in accordance with the $U(1)_A$ anomaly. It provides a mechanical interpretation of the
chiral behavior of the topological susceptibility and illustrates the interplay of chiral
dynamics and topological fluctuations in low-energy QCD.

Gluon operators with axial vector quantum numbers appear in the operator expansion
of heavy flavor contributions to the nucleon spin structure functions \cite{Franz:2000ee};
their nucleon matrix elements can be evaluated using similar methods as described here.

\section{Beyond instantons}
\label{sec:beyond}
The instanton vacuum describes the nonperturbative QCD gauge fields as a superposition
of well-separated instantons. The instanton fields are special in that they carry local
topological charge $\pm 1$ and induce zero modes of the fermion fields, and thus cause ChSB,
which determines the effective dynamics and the structure of light hadrons.
This review covers the distinct gluonic structure induced by well-separated instantons,
relevant for the reasons summarized in Sec.~\ref{sec:intro}.
Other vacuum fluctuations can also give rise to strong gauge fields and
make significant contributions to the gluonic structure of hadrons.

A broader view of the possible vacuum fluctuations can be obtained from an analysis
of the topological landscape of Yang-Mills gauge theory. Gauge fields configurations
are characterized by a winding number $N_{\rm CS}$ (Chern-Simons number);
the energy is a periodic function
of $N_{\rm CS}$, with minima at integer values and a finite potential barrier between them;
the height of barrier is determined by the size of the field configurations, $\rho$
\cite{Schafer:1996wv,Diakonov:2009jq}. Instantons represent
semiclassical tunneling trajectories between minima with $\Delta N_{\rm CS} = \pm 1$ at zero energy.
Recent work \cite{Shuryak:2021fsu} has explored the effects of other semiclassical trajectories in the topological
landscape: (i)~instanton-antiinstanton molecules, or streamline paths, where the semiclassical
motion does not result in tunneling (``failed tunneling trajectories'').
(ii)~finite-energy tunneling trajectories, or zig-zag paths, where the tunneling process occurs
at finite energy and includes both real-time and imaginary-time motion.
These types of field configurations are different from well-separated instantons;
they do not induce fermion zero modes and thus do not contribute to ChSB.
However, they give rise to strong fields and contribute to Wilson loops \cite{Schafer:1996wv},
high-momentum transfer processes (form factors), and other structures.
They should therefore be included in the semiclassical description of QCD vacuum.

Including the instanton-antiinstanton molecules in the variational description of instanton
vacuum poses several questions \cite{Shuryak:2021fsu}.
One needs to reassess the instanton density (especially
at small sizes $\rho < \bar\rho$) and revise the notion of instanton interaction.
One also needs to revisit the instanton packing fraction expansion, which provides
a simple ordering scheme and guarantees conservation laws
(current conservation, energy-momentum conservation).
The benefit of including instanton-antiinstanton molecules is that one obtains
quantitative estimates for the many structures that are absent in the dilute instanton medium,
especially the twist-2 gluon density.

Including the effects of instanton-antiinstanton molecules in the effective dynamics arising from ChSB
($1/N_c$ expansion, chiral soliton picture of nucleon) and in the effective operator approach
presents many opportunities for further development. It can lead to a more realistic
semiclassical description of light hadron structure, especially gluonic structure \cite{Liu:2024rdm}.

\textit{Related subjects.}
This review focuses on the gluonic properties of light hadrons, whose structure is
essentially determined by ChSB. The instanton vacuum can also be employed to study heavy-light
hadrons \cite{Chernyshev:1994zm,Chernyshev:1995gj,Hong:2024ptu}, where ChSB controls the
dynamics of the light flavors in the presence of the heavy quark. The instanton vacuum and its
extensions can also be applied to study heavy
quarkonia \cite{Diakonov:1989un,Yakhshiev:2016keg,Musakhanov:2020hvk,Shuryak:2021yif,Miesch:2023hjt,Miesch:2024fhv};
in these systems semiclassical trajectories beyond well-separated instantons are expected to play
an important role.

The effective operator method can also be used to compute vacuum expectation values of
higher-dimensional QCD quark-gluon operators, such as the dimension-5 chiral-odd operator
$\bar\psi (x) (\lambda^a/2) \sigma^{\mu\nu} \psi (x) F^a_{\mu\nu}(x)$ [of the form Eq.~(\ref{operator_color_octet})]
and similar dimension-7 operators \cite{Polyakov:1996kh,Polyakov:1998ip}. These higher-dimensional
vacuum condensates are generalizations of the chiral order parameter and sensitive to
the gluon fields active in ChSB. Comparison of the instanton vacuum results with
lattice QCD calculations provides direct information on the instanton size distribution
in the vacuum, complementary to cooling studies.

\textit{Acknowledgments.}
This material is based upon work supported by the U.S.~Department of Energy, Office of Science,
Office of Nuclear Physics under contract DE-AC05-06OR23177. 
The research reported here takes place in the context of the Topical Collaboration ``3D quark-gluon
structure of hadrons: mass, spin, tomography'' (Quark-Gluon Tomography Collaboration) supported by
the U.S.~Department of Energy, Office of Science, Office of Nuclear Physics under
contract DE-SC0023646.
\bibliography{memorial_instanton}

\begin{thebibliography}{10}
\expandafter\ifx\csname url\endcsname\relax
  \def\url#1{\texttt{#1}}\fi
\expandafter\ifx\csname urlprefix\endcsname\relax\def\urlprefix{URL }\fi
\expandafter\ifx\csname href\endcsname\relax
  \def\href#1#2{#2} \def\path#1{#1}\fi

\bibitem{Callan:1976je}
C.~G. Callan, Jr., R.~F. Dashen, D.~J. Gross, {The Structure of the Gauge
  Theory Vacuum}, Phys. Lett. B 63 (1976) 334--340.
\newblock \href {http://dx.doi.org/10.1016/0370-2693(76)90277-X}
  {\path{doi:10.1016/0370-2693(76)90277-X}}.

\bibitem{Callan:1977gz}
C.~G. Callan, Jr., R.~F. Dashen, D.~J. Gross, {Toward a Theory of the Strong
  Interactions}, Phys. Rev. D 17 (1978) 2717.
\newblock \href {http://dx.doi.org/10.1103/PhysRevD.17.2717}
  {\path{doi:10.1103/PhysRevD.17.2717}}.

\bibitem{Vainshtein:1981wh}
A.~I. Vainshtein, V.~I. Zakharov, V.~A. Novikov, M.~A. Shifman, {ABC's of
  Instantons}, Sov. Phys. Usp. 25 (1982) 195.
\newblock \href {http://dx.doi.org/10.1070/PU1982v025n04ABEH004533}
  {\path{doi:10.1070/PU1982v025n04ABEH004533}}.

\bibitem{Banks:1979yr}
T.~Banks, A.~Casher, {Chiral Symmetry Breaking in Confining Theories}, Nucl.
  Phys. B 169 (1980) 103--125.
\newblock \href {http://dx.doi.org/10.1016/0550-3213(80)90255-2}
  {\path{doi:10.1016/0550-3213(80)90255-2}}.

\bibitem{Leinweber:1999cw}
D.~B. Leinweber, {Visualizations of the QCD vacuum}, in: {Workshop on
  Light-Cone QCD and Nonperturbative Hadron Physics}, 1999, pp. 138--143.
\newblock \href {http://arxiv.org/abs/hep-lat/0004025}
  {\path{arXiv:hep-lat/0004025}}.

\bibitem{Bonati:2014tqa}
C.~Bonati, M.~D'Elia, {Comparison of the gradient flow with cooling in $SU(3)$
  pure gauge theory}, Phys. Rev. D 89~(10) (2014) 105005.
\newblock \href {http://arxiv.org/abs/1401.2441} {\path{arXiv:1401.2441}},
  \href {http://dx.doi.org/10.1103/PhysRevD.89.105005}
  {\path{doi:10.1103/PhysRevD.89.105005}}.

\bibitem{Alexandrou:2017hqw}
C.~Alexandrou, A.~Athenodorou, K.~Cichy, A.~Dromard, E.~Garcia-Ramos,
  K.~Jansen, U.~Wenger, F.~Zimmermann, {Comparison of topological charge
  definitions in Lattice QCD}, Eur. Phys. J. C 80~(5) (2020) 424.
\newblock \href {http://arxiv.org/abs/1708.00696} {\path{arXiv:1708.00696}},
  \href {http://dx.doi.org/10.1140/epjc/s10052-020-7984-9}
  {\path{doi:10.1140/epjc/s10052-020-7984-9}}.

\bibitem{Athenodorou:2018jwu}
A.~Athenodorou, P.~Boucaud, F.~De~Soto, J.~Rodr\'\i{}guez-Quintero,
  S.~Zafeiropoulos, {Instanton liquid properties from lattice QCD}, JHEP 02
  (2018) 140.
\newblock \href {http://arxiv.org/abs/1801.10155} {\path{arXiv:1801.10155}},
  \href {http://dx.doi.org/10.1007/JHEP02(2018)140}
  {\path{doi:10.1007/JHEP02(2018)140}}.

\bibitem{Shuryak:1981ff}
E.~V. Shuryak, {The Role of Instantons in Quantum Chromodynamics. 1. Physical
  Vacuum}, Nucl. Phys. B 203 (1982) 93.
\newblock \href {http://dx.doi.org/10.1016/0550-3213(82)90478-3}
  {\path{doi:10.1016/0550-3213(82)90478-3}}.

\bibitem{Diakonov:1983hh}
D.~Diakonov, V.~Y. Petrov, {Instanton Based Vacuum from Feynman Variational
  Principle}, Nucl. Phys. B 245 (1984) 259--292.
\newblock \href {http://dx.doi.org/10.1016/0550-3213(84)90432-2}
  {\path{doi:10.1016/0550-3213(84)90432-2}}.

\bibitem{Diakonov:2002fq}
D.~Diakonov, {Instantons at work}, Prog. Part. Nucl. Phys. 51 (2003) 173--222.
\newblock \href {http://arxiv.org/abs/hep-ph/0212026}
  {\path{arXiv:hep-ph/0212026}}, \href
  {http://dx.doi.org/10.1016/S0146-6410(03)90014-7}
  {\path{doi:10.1016/S0146-6410(03)90014-7}}.

\bibitem{Schafer:1996wv}
T.~Sch\"afer, E.~V. Shuryak, {Instantons in QCD}, Rev. Mod. Phys. 70 (1998)
  323--426.
\newblock \href {http://arxiv.org/abs/hep-ph/9610451}
  {\path{arXiv:hep-ph/9610451}}, \href
  {http://dx.doi.org/10.1103/RevModPhys.70.323}
  {\path{doi:10.1103/RevModPhys.70.323}}.

\bibitem{Shuryak:1982dp}
E.~V. Shuryak, {The Role of Instantons in Quantum Chromodynamics. 2. Hadronic
  Structure}, Nucl. Phys. B 203 (1982) 116--139.
\newblock \href {http://dx.doi.org/10.1016/0550-3213(82)90479-5}
  {\path{doi:10.1016/0550-3213(82)90479-5}}.

\bibitem{Diakonov:1985eg}
D.~Diakonov, V.~Y. Petrov, {A Theory of Light Quarks in the Instanton Vacuum},
  Nucl. Phys. B 272 (1986) 457--489.
\newblock \href {http://dx.doi.org/10.1016/0550-3213(86)90011-8}
  {\path{doi:10.1016/0550-3213(86)90011-8}}.

\bibitem{Diakonov:1986aj}
D.~Diakonov, V.~Y. Petrov, {Spontaneous breaking of chiral symmetry in the
  instanton vacuum}, LENINGRAD-86-1153. Published (in Russian) in: Hadron
  Matter under Extreme Conditions, Eds.\ G.~M.~Zinovev and V.~P.~Shelest,
  Naukova Dumka, Kiev, p. 192., 1986.

\bibitem{Pobylitsa:1989uq}
P.~V. Pobylitsa, {The Quark Propagator and Correlation Functions in the
  Instanton Vacuum}, Phys. Lett. B 226 (1989) 387--392.
\newblock \href {http://dx.doi.org/10.1016/0370-2693(89)91216-1}
  {\path{doi:10.1016/0370-2693(89)91216-1}}.

\bibitem{Nowak:1989at}
M.~A. Nowak, J.~J.~M. Verbaarschot, I.~Zahed, {Instantons and Chiral Dynamics},
  Phys. Lett. B 228 (1989) 251--258.
\newblock \href {http://dx.doi.org/10.1016/0370-2693(89)90667-9}
  {\path{doi:10.1016/0370-2693(89)90667-9}}.

\bibitem{Diakonov:1995qy}
D.~Diakonov, M.~V. Polyakov, C.~Weiss, {Hadronic matrix elements of gluon
  operators in the instanton vacuum}, Nucl. Phys. B 461 (1996) 539--580.
\newblock \href {http://arxiv.org/abs/hep-ph/9510232}
  {\path{arXiv:hep-ph/9510232}}, \href
  {http://dx.doi.org/10.1016/0550-3213(95)00675-3}
  {\path{doi:10.1016/0550-3213(95)00675-3}}.

\bibitem{Kacir:1996qn}
M.~Kacir, M.~Prakash, I.~Zahed, {Hadrons and QCD instantons: A Bosonized view},
  Acta Phys. Polon. B 30 (1999) 287--348.
\newblock \href {http://arxiv.org/abs/hep-ph/9602314}
  {\path{arXiv:hep-ph/9602314}}.

\bibitem{Diakonov:1987ty}
D.~Diakonov, V.~Y. Petrov, P.~V. Pobylitsa, {A Chiral Theory of Nucleons},
  Nucl. Phys. B 306 (1988) 809.
\newblock \href {http://dx.doi.org/10.1016/0550-3213(88)90443-9}
  {\path{doi:10.1016/0550-3213(88)90443-9}}.

\bibitem{Witten:1979kh}
E.~Witten, {Baryons in the $1/N$ Expansion}, Nucl. Phys. B 160 (1979) 57--115.
\newblock \href {http://dx.doi.org/10.1016/0550-3213(79)90232-3}
  {\path{doi:10.1016/0550-3213(79)90232-3}}.

\bibitem{Christov:1995vm}
C.~V. Christov, A.~Blotz, H.-C. Kim, P.~Pobylitsa, T.~Watabe, T.~Meissner,
  E.~Ruiz~Arriola, K.~Goeke, {Baryons as nontopological chiral solitons}, Prog.
  Part. Nucl. Phys. 37 (1996) 91--191.
\newblock \href {http://arxiv.org/abs/hep-ph/9604441}
  {\path{arXiv:hep-ph/9604441}}, \href
  {http://dx.doi.org/10.1016/0146-6410(96)00057-9}
  {\path{doi:10.1016/0146-6410(96)00057-9}}.

\bibitem{Liu:2024rdm}
W.-Y. Liu, E.~Shuryak, I.~Zahed, {Glue in hadrons at medium resolution and the
  QCD instanton vacuum}, Phys. Rev. D 110~(5) (2024) 054005.
\newblock \href {http://arxiv.org/abs/2404.03047} {\path{arXiv:2404.03047}},
  \href {http://dx.doi.org/10.1103/PhysRevD.110.054005}
  {\path{doi:10.1103/PhysRevD.110.054005}}.

\bibitem{tHooft:1976snw}
G.~'t~Hooft, {Computation of the Quantum Effects Due to a Four-Dimensional
  Pseudoparticle}, Phys. Rev. D 14 (1976) 3432--3450, [Erratum: Phys.Rev.D 18,
  2199 (1978)].
\newblock \href {http://dx.doi.org/10.1103/PhysRevD.14.3432}
  {\path{doi:10.1103/PhysRevD.14.3432}}.

\bibitem{Diakonov:2009jq}
D.~Diakonov, {Topology and confinement}, Nucl. Phys. B Proc. Suppl. 195 (2009)
  5--45.
\newblock \href {http://arxiv.org/abs/0906.2456} {\path{arXiv:0906.2456}},
  \href {http://dx.doi.org/10.1016/j.nuclphysbps.2009.10.010}
  {\path{doi:10.1016/j.nuclphysbps.2009.10.010}}.

\bibitem{Diakonov:1983bny}
D.~Diakonov, M.~I. Eides, {Chiral lagrangian from a functional integral over
  quarks}, JETP Lett. 38 (1983) 433--436.

\bibitem{Balla:1997hf}
J.~Balla, M.~V. Polyakov, C.~Weiss, {Nucleon matrix elements of higher twist
  operators from the instanton vacuum}, Nucl. Phys. B 510 (1998) 327--364.
\newblock \href {http://arxiv.org/abs/hep-ph/9707515}
  {\path{arXiv:hep-ph/9707515}}, \href
  {http://dx.doi.org/10.1016/S0550-3213(98)00638-5}
  {\path{doi:10.1016/S0550-3213(98)00638-5}}.

\bibitem{Weiss:2021kpt}
C.~Weiss, {Nucleon matrix element of Weinberg's CP-odd gluon operator from the
  instanton vacuum}, Phys. Lett. B 819 (2021) 136447.
\newblock \href {http://arxiv.org/abs/2103.13471} {\path{arXiv:2103.13471}},
  \href {http://dx.doi.org/10.1016/j.physletb.2021.136447}
  {\path{doi:10.1016/j.physletb.2021.136447}}.

\bibitem{Dressler:1999zi}
B.~Dressler, M.~Maul, C.~Weiss, {Twist four contribution to unpolarized
  structure functions $F_L$ and $F_2$ from instantons}, Nucl. Phys. B 578
  (2000) 293--325.
\newblock \href {http://arxiv.org/abs/hep-ph/9906444}
  {\path{arXiv:hep-ph/9906444}}, \href
  {http://dx.doi.org/10.1016/S0550-3213(00)00024-9}
  {\path{doi:10.1016/S0550-3213(00)00024-9}}.

\bibitem{Liu:2025ldh}
W.-Y. Liu, {Generic framework for non-perturbative QCD in light hadrons}\href
  {http://arxiv.org/abs/2501.07776} {\path{arXiv:2501.07776}}.

\bibitem{Kim:2023pll}
J.-Y. Kim, C.~Weiss, {Instanton effects in twist-3 generalized parton
  distributions}, Phys. Lett. B 848 (2024) 138387.
\newblock \href {http://arxiv.org/abs/2310.16890} {\path{arXiv:2310.16890}},
  \href {http://dx.doi.org/10.1016/j.physletb.2023.138387}
  {\path{doi:10.1016/j.physletb.2023.138387}}.

\bibitem{Diakonov:1996sr}
D.~Diakonov, V.~Petrov, P.~Pobylitsa, M.~V. Polyakov, C.~Weiss, {Nucleon parton
  distributions at low normalization point in the large $N_c$ limit}, Nucl.
  Phys. B 480 (1996) 341--380.
\newblock \href {http://arxiv.org/abs/hep-ph/9606314}
  {\path{arXiv:hep-ph/9606314}}, \href
  {http://dx.doi.org/10.1016/S0550-3213(96)00486-5}
  {\path{doi:10.1016/S0550-3213(96)00486-5}}.

\bibitem{Diakonov:1997vc}
D.~Diakonov, V.~Y. Petrov, P.~V. Pobylitsa, M.~V. Polyakov, C.~Weiss,
  {Unpolarized and polarized quark distributions in the large $N_c$ limit},
  Phys. Rev. D 56 (1997) 4069--4083.
\newblock \href {http://arxiv.org/abs/hep-ph/9703420}
  {\path{arXiv:hep-ph/9703420}}, \href
  {http://dx.doi.org/10.1103/PhysRevD.56.4069}
  {\path{doi:10.1103/PhysRevD.56.4069}}.

\bibitem{STAR:2014afm}
L.~Adamczyk, et~al., {Measurement of longitudinal spin asymmetries for weak
  boson production in polarized proton-proton collisions at RHIC}, Phys. Rev.
  Lett. 113 (2014) 072301.
\newblock \href {http://arxiv.org/abs/1404.6880} {\path{arXiv:1404.6880}},
  \href {http://dx.doi.org/10.1103/PhysRevLett.113.072301}
  {\path{doi:10.1103/PhysRevLett.113.072301}}.

\bibitem{STAR:2018fty}
J.~Adam, et~al., {Measurement of the longitudinal spin asymmetries for weak
  boson production in proton-proton collisions at $\sqrt{s}$ = 510 GeV}, Phys.
  Rev. D 99~(5) (2019) 051102.
\newblock \href {http://arxiv.org/abs/1812.04817} {\path{arXiv:1812.04817}},
  \href {http://dx.doi.org/10.1103/PhysRevD.99.051102}
  {\path{doi:10.1103/PhysRevD.99.051102}}.

\bibitem{PHENIX:2015ade}
A.~Adare, et~al., {Measurement of parity-violating spin asymmetries in
  W$^{\pm}$ production at midrapidity in longitudinally polarized $p+p$
  collisions}, Phys. Rev. D 93~(5) (2016) 051103.
\newblock \href {http://arxiv.org/abs/1504.07451} {\path{arXiv:1504.07451}},
  \href {http://dx.doi.org/10.1103/PhysRevD.93.051103}
  {\path{doi:10.1103/PhysRevD.93.051103}}.

\bibitem{PHENIX:2018wuz}
A.~Adare, et~al., {Cross section and longitudinal single-spin asymmetry $A_L$
  for forward $W^{\pm}\rightarrow\mu^{\pm}\nu$ production in polarized $p+p$
  collisions at $\sqrt{s}=510$ GeV}, Phys. Rev. D 98~(3) (2018) 032007.
\newblock \href {http://arxiv.org/abs/1804.04181} {\path{arXiv:1804.04181}},
  \href {http://dx.doi.org/10.1103/PhysRevD.98.032007}
  {\path{doi:10.1103/PhysRevD.98.032007}}.

\bibitem{Cocuzza:2022jye}
C.~Cocuzza, W.~Melnitchouk, A.~Metz, N.~Sato, {Polarized antimatter in the
  proton from a global QCD analysis}, Phys. Rev. D 106~(3) (2022) L031502.
\newblock \href {http://arxiv.org/abs/2202.03372} {\path{arXiv:2202.03372}},
  \href {http://dx.doi.org/10.1103/PhysRevD.106.L031502}
  {\path{doi:10.1103/PhysRevD.106.L031502}}.

\bibitem{HadStruc:2022nay}
R.~G. Edwards, et~al., {Non-singlet quark helicity PDFs of the nucleon from
  pseudo-distributions}, JHEP 03 (2023) 086.
\newblock \href {http://arxiv.org/abs/2211.04434} {\path{arXiv:2211.04434}},
  \href {http://dx.doi.org/10.1007/JHEP03(2023)086}
  {\path{doi:10.1007/JHEP03(2023)086}}.

\bibitem{Shuryak:2021fsu}
E.~Shuryak, I.~Zahed, {Hadronic structure on the light front. I. Instanton
  effects and quark-antiquark effective potentials}, Phys. Rev. D 107~(3)
  (2023) 034023.
\newblock \href {http://arxiv.org/abs/2110.15927} {\path{arXiv:2110.15927}},
  \href {http://dx.doi.org/10.1103/PhysRevD.107.034023}
  {\path{doi:10.1103/PhysRevD.107.034023}}.

\bibitem{Shuryak:2021hng}
E.~Shuryak, I.~Zahed, {Hadronic structure on the light front. II. QCD strings,
  Wilson lines, and potentials}, Phys. Rev. D 107~(3) (2023) 034024.
\newblock \href {http://arxiv.org/abs/2111.01775} {\path{arXiv:2111.01775}},
  \href {http://dx.doi.org/10.1103/PhysRevD.107.034024}
  {\path{doi:10.1103/PhysRevD.107.034024}}.

\bibitem{Liu:2024jno}
W.-Y. Liu, E.~Shuryak, C.~Weiss, I.~Zahed, {Pion gravitational form factors in
  the QCD instanton vacuum. I}, Phys. Rev. D 110~(5) (2024) 054021.
\newblock \href {http://arxiv.org/abs/2405.14026} {\path{arXiv:2405.14026}},
  \href {http://dx.doi.org/10.1103/PhysRevD.110.054021}
  {\path{doi:10.1103/PhysRevD.110.054021}}.

\bibitem{Gluck:1998xa}
M.~Gl\"uck, E.~Reya, A.~Vogt, {Dynamical parton distributions revisited}, Eur.
  Phys. J. C 5 (1998) 461--470.
\newblock \href {http://arxiv.org/abs/hep-ph/9806404}
  {\path{arXiv:hep-ph/9806404}}, \href
  {http://dx.doi.org/10.1007/s100520050289} {\path{doi:10.1007/s100520050289}}.

\bibitem{Lorce:2017wkb}
C.~Lorc\'e, L.~Mantovani, B.~Pasquini, {Spatial distribution of angular
  momentum inside the nucleon}, Phys. Lett. B 776 (2018) 38--47.
\newblock \href {http://arxiv.org/abs/1704.08557} {\path{arXiv:1704.08557}},
  \href {http://dx.doi.org/10.1016/j.physletb.2017.11.018}
  {\path{doi:10.1016/j.physletb.2017.11.018}}.

\bibitem{Kim:2024cbq}
J.-Y. Kim, H.-Y. Won, H.-C. Kim, C.~Weiss, {Spin-orbit correlations in the
  nucleon in the large-$N_c$ limit}, Phys. Rev. D 110~(5) (2024) 054026.
\newblock \href {http://arxiv.org/abs/2403.07186} {\path{arXiv:2403.07186}},
  \href {http://dx.doi.org/10.1103/PhysRevD.110.054026}
  {\path{doi:10.1103/PhysRevD.110.054026}}.

\bibitem{Lorce:2014mxa}
C.~Lorc\'e, {Spin\textendash{}orbit correlations in the nucleon}, Phys. Lett. B
  735 (2014) 344--348.
\newblock \href {http://arxiv.org/abs/1401.7784} {\path{arXiv:1401.7784}},
  \href {http://dx.doi.org/10.1016/j.physletb.2014.06.068}
  {\path{doi:10.1016/j.physletb.2014.06.068}}.

\bibitem{Shuryak:1981pi}
E.~V. Shuryak, A.~I. Vainshtein, {Theory of Power Corrections to Deep Inelastic
  Scattering in Quantum Chromodynamics. 2. $Q^4$ Effects: Polarized Target},
  Nucl. Phys. B 201 (1982) 141.
\newblock \href {http://dx.doi.org/10.1016/0550-3213(82)90377-7}
  {\path{doi:10.1016/0550-3213(82)90377-7}}.

\bibitem{Shuryak:1981kj}
E.~V. Shuryak, A.~I. Vainshtein, {Theory of Power Corrections to Deep Inelastic
  Scattering in Quantum Chromodynamics. 1. $Q^2$ Effects}, Nucl. Phys. B 199
  (1982) 451--481.
\newblock \href {http://dx.doi.org/10.1016/0550-3213(82)90355-8}
  {\path{doi:10.1016/0550-3213(82)90355-8}}.

\bibitem{Kuhn:2008sy}
S.~E. Kuhn, J.~P. Chen, E.~Leader, {Spin Structure of the Nucleon - Status and
  Recent Results}, Prog. Part. Nucl. Phys. 63 (2009) 1--50.
\newblock \href {http://arxiv.org/abs/0812.3535} {\path{arXiv:0812.3535}},
  \href {http://dx.doi.org/10.1016/j.ppnp.2009.02.001}
  {\path{doi:10.1016/j.ppnp.2009.02.001}}.

\bibitem{Ehrnsperger:1993hh}
B.~Ehrnsperger, A.~Schafer, L.~Mankiewicz, {OPE analysis for polarized deep
  inelastic scattering}, Phys. Lett. B 323 (1994) 439--445.
\newblock \href {http://arxiv.org/abs/hep-ph/9311285}
  {\path{arXiv:hep-ph/9311285}}, \href
  {http://dx.doi.org/10.1016/0370-2693(94)91244-0}
  {\path{doi:10.1016/0370-2693(94)91244-0}}.

\bibitem{Sato:2016tuz}
N.~Sato, W.~Melnitchouk, S.~E. Kuhn, J.~J. Ethier, A.~Accardi, {Iterative Monte
  Carlo analysis of spin-dependent parton distributions}, Phys. Rev. D 93~(7)
  (2016) 074005.
\newblock \href {http://arxiv.org/abs/1601.07782} {\path{arXiv:1601.07782}},
  \href {http://dx.doi.org/10.1103/PhysRevD.93.074005}
  {\path{doi:10.1103/PhysRevD.93.074005}}.

\bibitem{Gockeler:2005vw}
M.~Gockeler, R.~Horsley, D.~Pleiter, P.~E.~L. Rakow, A.~Schafer, G.~Schierholz,
  H.~Stuben, J.~M. Zanotti, {Investigation of the second moment of the
  nucleon's $g_1$ and $g_2$ structure functions in two-flavor lattice QCD},
  Phys. Rev. D 72 (2005) 054507.
\newblock \href {http://arxiv.org/abs/hep-lat/0506017}
  {\path{arXiv:hep-lat/0506017}}, \href
  {http://dx.doi.org/10.1103/PhysRevD.72.054507}
  {\path{doi:10.1103/PhysRevD.72.054507}}.

\bibitem{Sidorov:2006vu}
A.~V. Sidorov, C.~Weiss, {Higher twists in polarized DIS and the size of the
  constituent quark}, Phys. Rev. D 73 (2006) 074016.
\newblock \href {http://arxiv.org/abs/hep-ph/0602142}
  {\path{arXiv:hep-ph/0602142}}, \href
  {http://dx.doi.org/10.1103/PhysRevD.73.074016}
  {\path{doi:10.1103/PhysRevD.73.074016}}.

\bibitem{Lee:2001ug}
N.-Y. Lee, K.~Goeke, C.~Weiss, {Spin dependent twist four matrix elements from
  the instanton vacuum: Flavor singlet and nonsinglet}, Phys. Rev. D 65 (2002)
  054008.
\newblock \href {http://arxiv.org/abs/hep-ph/0105173}
  {\path{arXiv:hep-ph/0105173}}, \href
  {http://dx.doi.org/10.1103/PhysRevD.65.054008}
  {\path{doi:10.1103/PhysRevD.65.054008}}.

\bibitem{Weiss:2002qq}
C.~Weiss, {Modeling power corrections to the Bjorken sum rule for the neutrino
  structure function $F_1$}, J. Phys. G 29 (2003) 1981--1984.
\newblock \href {http://arxiv.org/abs/hep-ph/0210132}
  {\path{arXiv:hep-ph/0210132}}, \href
  {http://dx.doi.org/10.1088/0954-3899/29/8/385}
  {\path{doi:10.1088/0954-3899/29/8/385}}.

\bibitem{Dressler:1999hc}
B.~Dressler, M.~V. Polyakov, {On the twist-three contribution to $h_L$ in the
  instanton vacuum}, Phys. Rev. D 61 (2000) 097501.
\newblock \href {http://arxiv.org/abs/hep-ph/9912376}
  {\path{arXiv:hep-ph/9912376}}, \href
  {http://dx.doi.org/10.1103/PhysRevD.61.097501}
  {\path{doi:10.1103/PhysRevD.61.097501}}.

\bibitem{Kiptily:2002nx}
D.~V. Kiptily, M.~V. Polyakov, {Genuine twist three contributions to the
  generalized parton distributions from instantons}, Eur. Phys. J. C 37 (2004)
  105--114.
\newblock \href {http://arxiv.org/abs/hep-ph/0212372}
  {\path{arXiv:hep-ph/0212372}}, \href
  {http://dx.doi.org/10.1140/epjc/s2004-01957-3}
  {\path{doi:10.1140/epjc/s2004-01957-3}}.

\bibitem{Liu:2021evw}
Y.~Liu, I.~Zahed, {Small size instanton contributions to the quark quasi-PDF
  and matching kernel}\href {http://arxiv.org/abs/2102.07248}
  {\path{arXiv:2102.07248}}.

\bibitem{Novikov:1981xi}
V.~A. Novikov, M.~A. Shifman, A.~I. Vainshtein, V.~I. Zakharov, {Are All
  Hadrons Alike?}, Nucl. Phys. B 191 (1981) 301--369.
\newblock \href {http://dx.doi.org/10.1016/0550-3213(81)90303-5}
  {\path{doi:10.1016/0550-3213(81)90303-5}}.

\bibitem{Lorce:2018egm}
C.~Lorc\'e, H.~Moutarde, A.~P. Trawi\'nski, {Revisiting the mechanical
  properties of the nucleon}, Eur. Phys. J. C 79~(1) (2019) 89.
\newblock \href {http://arxiv.org/abs/1810.09837} {\path{arXiv:1810.09837}},
  \href {http://dx.doi.org/10.1140/epjc/s10052-019-6572-3}
  {\path{doi:10.1140/epjc/s10052-019-6572-3}}.

\bibitem{Polyakov:2018zvc}
M.~V. Polyakov, P.~Schweitzer, {Forces inside hadrons: pressure, surface
  tension, mechanical radius, and all that}, Int. J. Mod. Phys. A 33~(26)
  (2018) 1830025.
\newblock \href {http://arxiv.org/abs/1805.06596} {\path{arXiv:1805.06596}},
  \href {http://dx.doi.org/10.1142/S0217751X18300259}
  {\path{doi:10.1142/S0217751X18300259}}.

\bibitem{Duran:2022xag}
B.~Duran, et~al., {Determining the gluonic gravitational form factors of the
  proton}, Nature 615~(7954) (2023) 813--816.
\newblock \href {http://arxiv.org/abs/2207.05212} {\path{arXiv:2207.05212}},
  \href {http://dx.doi.org/10.1038/s41586-023-05730-4}
  {\path{doi:10.1038/s41586-023-05730-4}}.

\bibitem{Liu:2024vkj}
W.-Y. Liu, E.~Shuryak, I.~Zahed, {Pion gravitational form factors in the QCD
  instanton vacuum. II}, Phys. Rev. D 110~(5) (2024) 054022.
\newblock \href {http://arxiv.org/abs/2405.16269} {\path{arXiv:2405.16269}},
  \href {http://dx.doi.org/10.1103/PhysRevD.110.054022}
  {\path{doi:10.1103/PhysRevD.110.054022}}.

\bibitem{Wang:2024lrm}
B.~Wang, F.~He, G.~Wang, T.~Draper, J.~Liang, K.-F. Liu, Y.-B. Yang, {Trace
  anomaly form factors from lattice QCD}, Phys. Rev. D 109~(9) (2024) 094504.
\newblock \href {http://arxiv.org/abs/2401.05496} {\path{arXiv:2401.05496}},
  \href {http://dx.doi.org/10.1103/PhysRevD.109.094504}
  {\path{doi:10.1103/PhysRevD.109.094504}}.

\bibitem{Hackett:2023nkr}
D.~C. Hackett, P.~R. Oare, D.~A. Pefkou, P.~E. Shanahan, {Gravitational form
  factors of the pion from lattice QCD}, Phys. Rev. D 108~(11) (2023) 114504.
\newblock \href {http://arxiv.org/abs/2307.11707} {\path{arXiv:2307.11707}},
  \href {http://dx.doi.org/10.1103/PhysRevD.108.114504}
  {\path{doi:10.1103/PhysRevD.108.114504}}.

\bibitem{Polyakov:2018exb}
M.~V. Polyakov, H.-D. Son, {Nucleon gravitational form factors from instantons:
  forces between quark and gluon subsystems}, JHEP 09 (2018) 156.
\newblock \href {http://arxiv.org/abs/1808.00155} {\path{arXiv:1808.00155}},
  \href {http://dx.doi.org/10.1007/JHEP09(2018)156}
  {\path{doi:10.1007/JHEP09(2018)156}}.

\bibitem{Franz:2000ee}
M.~Franz, M.~V. Polyakov, K.~Goeke, {Heavy quark mass expansion and intrinsic
  charm in light hadrons}, Phys. Rev. D 62 (2000) 074024.
\newblock \href {http://arxiv.org/abs/hep-ph/0002240}
  {\path{arXiv:hep-ph/0002240}}, \href
  {http://dx.doi.org/10.1103/PhysRevD.62.074024}
  {\path{doi:10.1103/PhysRevD.62.074024}}.

\bibitem{Chernyshev:1994zm}
S.~Chernyshev, M.~A. Nowak, I.~Zahed, {Heavy mesons in a random instanton gas},
  Phys. Lett. B 350 (1995) 238--244.
\newblock \href {http://arxiv.org/abs/hep-ph/9409207}
  {\path{arXiv:hep-ph/9409207}}, \href
  {http://dx.doi.org/10.1016/0370-2693(95)00330-N}
  {\path{doi:10.1016/0370-2693(95)00330-N}}.

\bibitem{Chernyshev:1995gj}
S.~Chernyshev, M.~A. Nowak, I.~Zahed, {Heavy hadrons and QCD instantons}, Phys.
  Rev. D 53 (1996) 5176--5184.
\newblock \href {http://arxiv.org/abs/hep-ph/9510326}
  {\path{arXiv:hep-ph/9510326}}, \href
  {http://dx.doi.org/10.1103/PhysRevD.53.5176}
  {\path{doi:10.1103/PhysRevD.53.5176}}.

\bibitem{Hong:2024ptu}
K.-H. Hong, H.-C. Kim, M.~M. Musakhanov, N.~Rakhimov, {Heavy-light quark
  systems from the QCD instanton vacuum: $N_f=1$ light flavor case}, Phys. Rev.
  D 110~(11) (2024) 114044.
\newblock \href {http://arxiv.org/abs/2410.13279} {\path{arXiv:2410.13279}},
  \href {http://dx.doi.org/10.1103/PhysRevD.110.114044}
  {\path{doi:10.1103/PhysRevD.110.114044}}.

\bibitem{Diakonov:1989un}
D.~Diakonov, V.~Y. Petrov, P.~V. Pobylitsa, {The Wilson Loop and Heavy Quark
  Potential in the Instanton Vacuum}, Phys. Lett. B 226 (1989) 372--376.
\newblock \href {http://dx.doi.org/10.1016/0370-2693(89)91213-6}
  {\path{doi:10.1016/0370-2693(89)91213-6}}.

\bibitem{Yakhshiev:2016keg}
U.~T. Yakhshiev, H.-C. Kim, M.~M. Musakhanov, E.~Hiyama, B.~Turimov, {Instanton
  effects on the heavy-quark static potential}, Chin. Phys. C 41~(8) (2017)
  083102.
\newblock \href {http://arxiv.org/abs/1602.06074} {\path{arXiv:1602.06074}},
  \href {http://dx.doi.org/10.1088/1674-1137/41/8/083102}
  {\path{doi:10.1088/1674-1137/41/8/083102}}.

\bibitem{Musakhanov:2020hvk}
M.~Musakhanov, N.~Rakhimov, U.~T. Yakhshiev, {Heavy quark correlators in an
  instanton liquid model with perturbative corrections}, Phys. Rev. D 102~(7)
  (2020) 076022.
\newblock \href {http://arxiv.org/abs/2006.01545} {\path{arXiv:2006.01545}},
  \href {http://dx.doi.org/10.1103/PhysRevD.102.076022}
  {\path{doi:10.1103/PhysRevD.102.076022}}.

\bibitem{Shuryak:2021yif}
E.~Shuryak, I.~Zahed, {Hadronic structure on the light front. III. The
  Hamiltonian, heavy quarkonia, spin, and orbit mixing}, Phys. Rev. D 107~(3)
  (2023) 034025.
\newblock \href {http://arxiv.org/abs/2112.15586} {\path{arXiv:2112.15586}},
  \href {http://dx.doi.org/10.1103/PhysRevD.107.034025}
  {\path{doi:10.1103/PhysRevD.107.034025}}.

\bibitem{Miesch:2023hjt}
N.~Miesch, E.~Shuryak, I.~Zahed, {Baryons and tetraquarks using
  instanton-induced interactions}, Phys. Rev. D 109~(1) (2024) 014022.
\newblock \href {http://arxiv.org/abs/2308.05638} {\path{arXiv:2308.05638}},
  \href {http://dx.doi.org/10.1103/PhysRevD.109.014022}
  {\path{doi:10.1103/PhysRevD.109.014022}}.

\bibitem{Miesch:2024fhv}
N.~Miesch, E.~Shuryak, I.~Zahed, {Bridging hadronic and vacuum structure by
  heavy quarkonia}, Phys. Rev. D 111~(3) (2025) 034006.
\newblock \href {http://arxiv.org/abs/2403.18700} {\path{arXiv:2403.18700}},
  \href {http://dx.doi.org/10.1103/PhysRevD.111.034006}
  {\path{doi:10.1103/PhysRevD.111.034006}}.

\bibitem{Polyakov:1996kh}
M.~V. Polyakov, C.~Weiss, {Mixed quark-gluon condensate from instantons}, Phys.
  Lett. B 387 (1996) 841--847.
\newblock \href {http://arxiv.org/abs/hep-ph/9607244}
  {\path{arXiv:hep-ph/9607244}}, \href
  {http://dx.doi.org/10.1016/0370-2693(96)01098-2}
  {\path{doi:10.1016/0370-2693(96)01098-2}}.

\bibitem{Polyakov:1998ip}
M.~V. Polyakov, C.~Weiss, {Estimates of higher dimensional vacuum condensates
  from the instanton vacuum}, Phys. Rev. D 57 (1998) 4471--4474.
\newblock \href {http://arxiv.org/abs/hep-ph/9710534}
  {\path{arXiv:hep-ph/9710534}}, \href
  {http://dx.doi.org/10.1103/PhysRevD.57.4471}
  {\path{doi:10.1103/PhysRevD.57.4471}}.

\end{thebibliography}
\end{document}